\documentclass[fleqn,usenatbib]{mnras}
\usepackage{newtxtext,newtxmath}
\usepackage[T1]{fontenc}
\DeclareRobustCommand{\VAN}[3]{#2}
\let\VANthebibliography\thebibliography
\def\thebibliography{\DeclareRobustCommand{\VAN}[3]{##3}\VANthebibliography}

\usepackage{graphicx}	
\usepackage{amsmath}	
\usepackage{longtable}
\usepackage{lscape}
\usepackage{afterpage}
\usepackage{subcaption}
\usepackage{xcolor}
\usepackage{booktabs}

\title[Search for protostellar jets with UWISH2]{Search for protostellar jets with UWISH2 in the molecular cloud complexes Vulpecula and IRDC G53.2}

\author[Chauhan et al.]{
Manish Chauhan,$^{1*}$\thanks{E-mail: manish.chauhan@outlook.in}
Manash Samal,$^{2}$\thanks{E-mail: manash@prl.res.in}
Anandmayee Tej,$^{1}$\thanks{E-mail: tej@iist.ac.in}
Dirk Froebrich$^{3}$
\\
$^{1}$Indian Institute of Space Science and Technology, Thiruvananthapuram, India\\
$^{2}$Physical Research Laboratory, Ahmedabad, India\\
$^{3}$University of Kent, Canterbury, UK\\
$^{*}$currently working at Space Applications Centre, ISRO, Ahmedabad, India
}

\date{Accepted XXX. Received YYY; in original form ZZZ}

\pubyear{2024}

\begin{document}
\label{firstpage}
\pagerange{\pageref{firstpage}--\pageref{lastpage}}
\maketitle

\begin{abstract}
Jets and outflows are the early signposts of stellar birth. Using the UKIRT Wide Field Infrared Survey for H$_2$ (UWISH2) at 2.12 $\mu$m, 127 outflows are identified in molecular cloud complexes Vulpecula OB1 and IRDC G53.2 covering 12 square degrees of the Galactic plane. Using multi-wavelength datasets, from 1.2 to 70 $\mu$m, 79 young stellar objects (YSOs) are proposed as potential driving sources, where, $\sim$79\% are likely Class 0/I protostars, 17\% are Class II YSOs and the remaining 4\% are Class III YSOs. The outflows are characterized in terms of their length, flux, luminosity and knot-spacing. The identified outflows have a median lobe length of 0.22 pc and 0.17 pc for outflows in Vulpecula OB1 and IRDC G53.2, respectively. Our analysis, from the knot spacing, reveals a typical ejection frequency of $\sim$1.2 kyr suggesting an intermediate type between the FU-Ori and EX-Ori type of eruptions in both cloud complexes. Furthermore, the physical parameters of the driving sources are obtained by performing radiative transfer modelling to the observed spectral energy distributions (SEDs), which suggest that the outflows are driven by intermediate mass stars. Various observed trends between the outflow properties and the corresponding driving sources, and various interesting outflows and star forming sites, including sites of triggered star formation and protocluster forming clump with clusters of jets, are discussed. The obtained results and the identified jet-bearing protostellar sample will pave the way to understand many aspects of outflows with future high-resolution observations.

\end{abstract}

	


\section{Introduction}\label{Sec:Introduction}

Jets and outflows are ubiquitous in star forming regions and are important signposts of stellar birth. Outflows represent entrained ambient interstellar medium (ISM) surrounding high velocity ejections from YSOs, and thus, provide fossil record of the associated mass accretion process \citep{Hartmann_1998}. This picture is supported by numerous theoretical and observational studies spanning more than a century \citep[e.g.][]{Burnham_1890, Herbig_1950,Haro_1952,Snell_1980}.

The protostellar jets can have velocities of a few tens to a hundred kms$^{-1}$ \citep{Reipurth_2001,Lee_2020} which allow them to carve out parsec-scale cavities in protostellar envelopes. Thus, they broadly influence the star formation process and growth of protostars or protoclusters in a variety of ways \citep[see reviews by][]{Bally_2016,Anglada_2018,Ray_2021}. For example, observations and numerical simulations suggest that protostellar jets/outflows inject a significant amount of energy and momentum into their surroundings and have a profound impact on their host cores/clumps \citep[e.g.][]{Arce_2010,Plunkett_2013}, thus, influencing the star-formation efficiency therein \citep[e.g.][]{naka07, Machida_2013, Wang_2010, fede14, Offner_2014, Krumholz_2019}. Similarly, theoretical as well as simulation works suggest that short outbursts with high accretion rate can mitigate the long-standing "luminosity problem" \citep{Kenyon_1990,Offner_2011}. Since, accretion and ejection are strongly coupled, episodic accretion events can be indirectly detected by the episodic outflows they trigger \citep{Arce_2007,Vorobyov_2018,Rohde_2019}. Thus, the spacing and kinematics of outflow bullets (or knots) and their properties can provide insights into the underlying episodic protostellar accretion. Numerical simulations also suggest that presence of a companion and interaction with sibling stars can strongly impact the outflow axis and shapes \citep[e.g.][]{Falle_1993,Terquem_1999, Bate_2000}. Testing and/or confirming these predictions lies critically in finding and characterizing a statistically robust sample of jets/outflows of different structures, morphologies, and evolutionary status, and linking them to the properties, characteristics, and motions of their driving sources or cores.

Emission from jets/outflows can be probed over a wide range of wavelengths. Different frequencies trace the emission from different regions of the shock-excited ISM resulting from collision with jets \citep{Bally_2016}. Observation of optical emission associated with these jets/outflows, dubbed as Herbig-Haro objects, is limited mostly to nearby star forming regions or massive outflows owing to large extinction towards Galactic star-forming clouds. Thus, jets/outflows in the Galactic plane are best traced at longer wavelengths as the effects of extinction is significantly reduced compared to the optical domain \citep[e.g.][]{Stanke_2002,Davis_2010,har15,rei17,rei22}.

Large sample of jet-bearing YSOs, similar to \citet{Makin_2018}, are necessary to obtain an in-depth knowledge of jets/outflows and their role in star formation. The UWISH2 survey \citep{Froebrich_2011} is an excellent database to directly probe jets/outflows in the Galactic plane. This is an unbiased narrow-band survey of the first Galactic quadrant (10$^\circ$ < l < 65$^\circ$, |b| < +1.3$^\circ$) centered on the H$_2$ ro-vibrational 1-0S(1) transition at 2.12 $\mu$m which is an excellent tracer of shock-excited, hot ($\sim$~2000~K), and dense (n~$\geq$~10$^3$~cm$^{-3}$) molecular gas. The survey was further extended to cover the Auriga and Cygnus region \citep[see][ F15 hereafter]{Froebrich_2015}. A number of surveys in the H$_2$ 1-0S(1) line transition at 2.12 $\mu$m, using data from UWISH2, reveal a large population of molecular hydrogen emission-line objects (MHOs) or outflows associated with YSOs in Galactic star-forming regions \citep{Ioannidis_2012a,Makin_2018, Makin_2019}.

In this paper, we conduct an extensive study of jets/outflows in two star-forming complexes, namely the Vulpecula OB association (hereafter Vul OB1) and the InfraRed Dark Cloud located at Galactic coordinates (l$\sim$53.2$^\circ$, b$\sim$0.0$^\circ$) (hereafter IRDC G53.2). The analysis is carried out using UWISH2 data in combination with multi-wavelength data retrieved from various archives (discussed in Section \ref{Sec:Data}). The two target regions provide excellent opportunity to study star formation in massive ($\sim10^5~$M$_\odot$) molecular clouds in the Galactic plane \citep[see][]{Wang_2020,Kohno_2022}. 
A large number of H$_2$ emission features are identified in these regions (refer to Section~\ref{Sec:Analysis_Result} for more details). IRDC G53.2 is a 45 pc long filamentary cloud located at a distance of 1.7~kpc with a rich YSO population of 373 candidate YSOs \citep{Kim_2015}. Vul OB1 ($l$ $\sim$ 60.2$^\circ$, $b$ $\sim$ -0.2$^\circ$) is an active star-forming site located at a distance of 2.3 kpc with around 856 YSO candidates identified by \citet{Billot_2010}. Vul OB1 also hosts the young open cluster NGC 6823 with an estimated age of 4$\pm$2 Myr \citep{Bica_2008}. Furthermore, three young HII regions are also seen to be associated with this region \citep[Sh2-86, Sh2-87 and Sh2-88;][]{Sharpless_1959}. Vul OB1 is affected by strong feedback from nearly 100 massive OB stars found in the region \citep{Reed_2003}. As a result, several feedback sculpted structures such as pillars or elephant-trunk like features have been identified in the complex \citep[e.g.][]{Billot_2010}. These pillars are important sites for studying triggered star formation \citep[e.g.][]{Billot_2010,Panwar_2019}, and detection of young outflows can possibly unravel induced star formation in such regions. 

In presenting the study, we structure the paper as follows. The data sets used in the analysis are discussed in Section \ref{Sec:Data}. The identification of MHOs, the likely driving source candidates and the distribution of various physical parameters of the MHOs and their driving sources are detailed in Section \ref{Sec:Analysis_Result}. Discussion on the correlations between the properties of the MHOs with the corresponding physical properties of the driving source candidates is given in Section \ref{Sec:Discussion}. In Section \ref{Sec:Notes}, we discuss a few interesting star-forming sites identified in this study. The results of the study are summarized in Section \ref{Sec:Conclusion}.

\section{Data sets}\label{Sec:Data}

\subsection{NIR data from UWISH2 and UKIDSS}

The UWISH2 survey mapped the First Galactic Quadrant of the Galactic plane using the Wide-Field Camera (WFCAM) at the United Kingdom Infrared Telescope (UKIRT) providing narrow band images centered on the H$_2$ 1-0S(1) transition. The survey has a 5$\sigma$ detection limit of 18 magnitudes in K-band and surface brightness limit of $\rm 10^{-19}\,Wm^{-2}arcsec^{-2}$ \citep{Froebrich_2011}. Continuum-subtracted H$_2$ line images retrieved from the UWISH2 archive\footnote{\url{http://astro.kent.ac.uk/uwish2/}} are used for the identification of jets/outflows. These images are obtained by subtracting the scaled K-band continuum images from the H$_2$ narrow-band images.

H$_2$ line emission from jets/knots appear as positive features in the continuum-subtracted images. Several
discrete jets/knots in a region can be associated to a coherent outflow. In this work, 
we use the discrete H$_2$ line-emission source catalogue for shock-excited jets by \citet{Froebrich_2015} to identify outflows of the studied complexes. We also utilize near-infrared (J, H and K band) photometric data and images taken as part of the UKIRT Infrared Deep Sky Survey's Galactic Plane Survey \citep[UKIDSS-GPS,][]{Lucas_2008} retrieved from the WFCAM Science Archive\footnote{\textcolor{blue}{\url{http://wsa.roe.ac.uk/}}}. J, H, and K-band photometric data are used for construction of extinction maps with the PNICER algorithm (refer Section \ref{PNICER}).

\subsection{CO data}

CO position-velocity data cubes from \citet{Dame_2001}\footnote{\url{https://www.cfa.harvard.edu/rtdc/CO/CompositeSurveys/}} and \citet{Jackson_2006}\footnote{\url{https://www.bu.edu/galacticring/new_data.html}} have been used to study the structure and extent of the Vul OB1 and IRDC G53.2 complexes, respectively. For IRDC G53.2, we use the $^{13}$CO (\textit{J}=1$\rightarrow$0) transition data from the Galactic Ring Survey \citep[GRS,][]{Jackson_2006} with a spatial resolution of 46\arcsec and velocity resolution of 0.21 kms$^{-1}$. In case of Vul OB1 complex, lower resolution (angular resolution of $\sim$ 7.5\arcmin and velocity resolution of 0.65~kms$^{-1}$) CO data cubes from \citet{Dame_2001}\footnote{\url{https://www.cfa.harvard.edu/rtdc/CO/}} have been used as the cloud falls outside the coverage of GRS (18$^\circ$ < $l$ < 55.7$^\circ$ and |$b$| < 1$^\circ$). The CO data is primarily used to define the outer extent of these clouds based on the integrated CO-emissions.

\subsection{Mid-Infrared data from {\it Spitzer} Space Telescope}

Mid-Infrared (MIR) data from the Galactic Legacy Infrared Midplane Survey Extraordinaire \citep[GLIMPSE,][]{Benjamin_2003,Churchwell_2009} in four IRAC bands between 3.6 and 8.0 $\mu$m and the MIPS GALactic plane survey at 24 $\mu$m \citep[MIPSGAL,][]{Carey_2009} of the {\it Spitzer} Space Telescope have been used for identification of driving source candidates for the identified jets/outflows. We used the Infrared Science Archive’s (IRSA) cutout service to obtain archival data for GLIMPSE and MIPSGAL surveys\footnote{\url{https://irsa.ipac.caltech.edu/applications/Cutouts/}}. The data products offer a spatial resolution of 2\arcsec and 6\arcsec for the {\it Spitzer} IRAC and MIPS band, respectively. 

\subsection{Far-Infrared data from Hi-GAL survey}

Far-Infrared (FIR) data from the $Herschel$ Infrared Galactic Plane Survey \citep[Hi-GAL,][]{Molinari_2010} are used in this study. The datasets retrieved from the science archives of the {\it Herschel} Space Observatory\footnote{\url{http://archives.esac.esa.int/hsa/whsa/}} include the Highly Processed Data Product (HPDP) images from the Photodetector Array Camera and Spectrometer (PACS; 70 $\mu$m and 160 $\mu$m) and the Spectral and Photometric Imaging Receiver (SPIRE; 250 $\mu$m, 350 $\mu$m, and 500 $\mu$m). The images have spatial resolutions of 6\arcsec, 12\arcsec, 17\arcsec, 24\arcsec, and 35\arcsec at 70, 160, 250, 350, and 500 $\mu$m, respectively. FIR data are used to identify deeply embedded YSO candidates and to investigate the nature of the driving sources. 

\subsection{Additional Data}
We also use images from the APEX Telescope Large Area Survey of the Galaxy \citep[ATLASGAL,][]{Schuller_2009} and the compact source catalogue of \citet{Csengeri_2014} at 870 $\mu$m, to examine the cold dust emission and identify clumps/cores in the vicinity of jets/knots. 
The 870 $\mu$m images, with a spatial resolution of 19.2\arcsec, are downloaded for the two complexes under study from the ATLASGAL Database Server\footnote{\textcolor{blue}{\url{http://atlasgal.mpifr-bonn.mpg.de/cgi-bin/ATLASGAL_DATABASE.cgi}}}. It should be noted that this catalogue, however, is limited to clumps with masses greater than 650 M$_\odot$. In addition, we have also examined the SiO emission catalogue of \citet{Csengeri_2016} around clumps. SiO emission is believed to originate from shock-excitation and can trace outflows from the youngest Class 0/I sources \citep[see][and references therein]{Samal_2018}. 

\section{Analysis and Results}\label{Sec:Analysis_Result}

\subsection{Identification of jets/outflows and driving source candidates}

\begin{figure}
  \centering
  \includegraphics[width=0.485\textwidth]{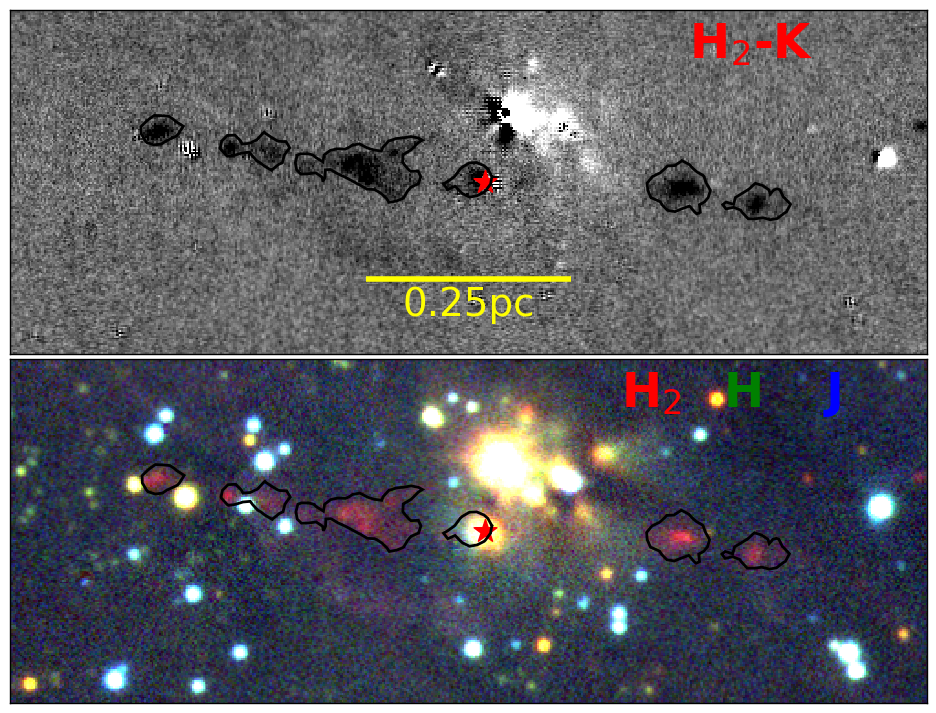}
  \caption{Example of a jet identified in Vul OB1. The top image is the continuum-subtracted (H$_2$-K) image from UWISH2 data and bottom panel shows the JHH$_2$ colour-composite image. The black contours shown are the extended H$_2$ line-emission sources from F15 catalogue. The location of the driving source is marked by a red star.}
  \label{fig:sample_jet}
\end{figure}

Studies of MHOs using UWISH2 data have revealed that majority of the regions observed by UWISH2 are devoid of outflows, $\sim$~2/3rd of the survey field in some cases \citep[e.g.][]{Ioannidis_2012a}. Keeping this in mind, we retrieve only the continuum-subtracted image tiles from the UWISH2 website with identified discrete H$_2$ line-emission sources from the F15 catalogue. In order to avoid any contamination or artefacts from bright stars, fluorescent emission, and other shock-excited features, we use only the features labelled as `jets' in the F15 catalogue. Jets/outflows from YSOs appear as chains of H$_2$ emission knots. The nature of these knots will be discussed in a later section (Section \ref{knot_gaps}). Figure \ref{fig:sample_jet} is an exemplar of aligned discrete H$_2$ knots in a star-forming site in Vul OB1, seen in the H$_2$-K image. The H$_2$ 1-0S(1) line-emission also manifests as red features in the JHH$_2$ colour-composite images, as seen in the lower panel of Figure \ref{fig:sample_jet}. The JHH$_2$ colour-composite images are used to confirm the classification of the H$_2$ features as jets and any possible contamination is removed.

Chains of H$_2$ features are linked to a single outflow based on the proximity and alignment of these knots. Following the scheme proposed by \citet{Davis_2010}, all morphologically well-connected H$_2$ emission objects are considered as a single MHO or outflow. The remaining discrete and isolated H$_2$ features are also considered as unique MHOs. As an example, in Figure \ref{fig:sample_jet}, all the six discrete emission features are considered to be part of a single MHO in our study. Using this approach, we identified a total of 127 MHOs in our study region, with Vul OB1 hosting 39\% of the identified outflows and IRDC G53.2 having 61\% of the outflows.  Detailed discussion on individual MHOs is given in the online material. While velocity information of individual knots are required to confirm their association, the morphology seen gives a strong indication that in all likelihood the proposed association would be correct.

\begin{figure*}
  \centering
  \includegraphics[width=0.85\textwidth]{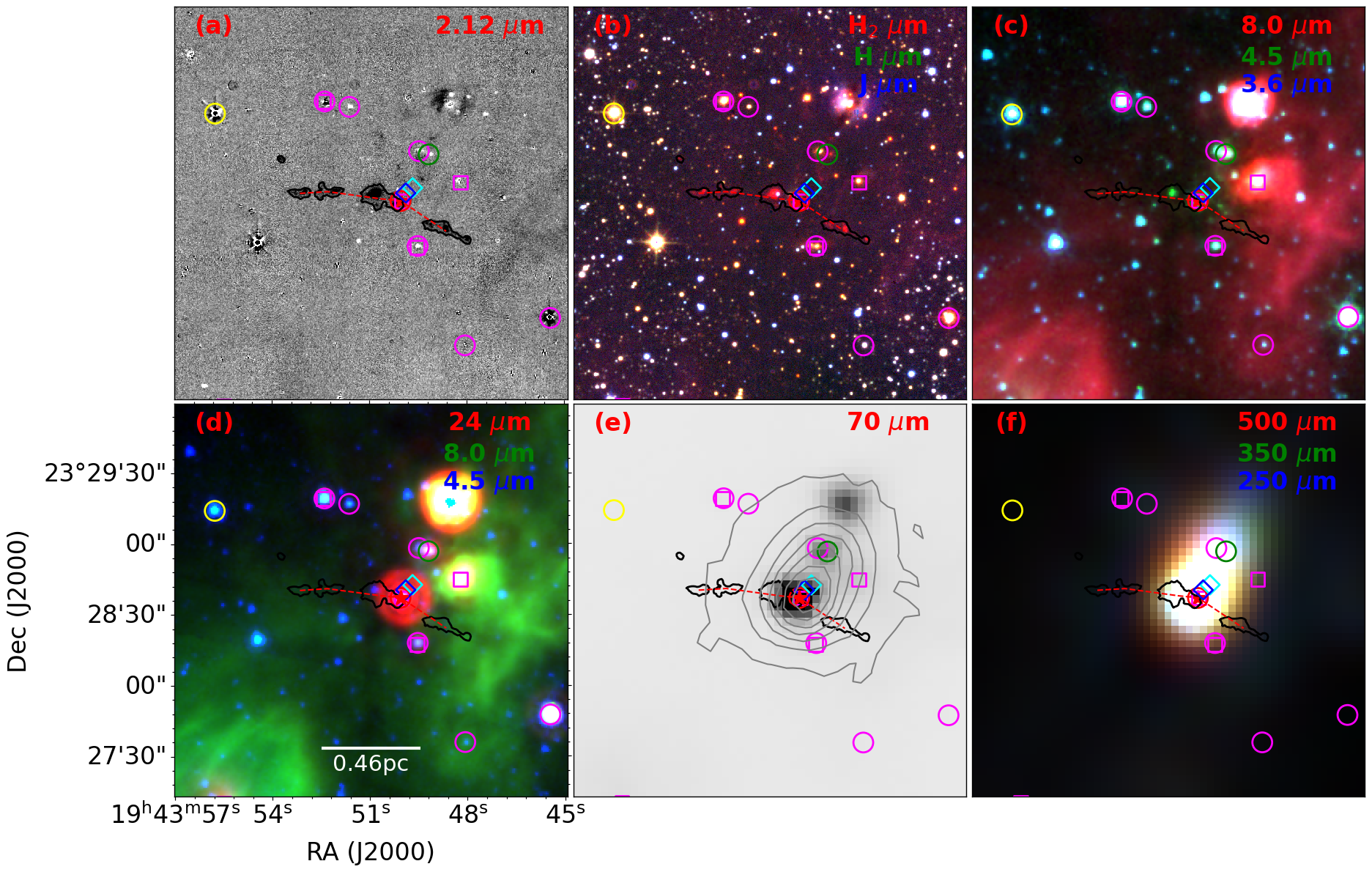}
  \caption{Colour-composite images for a curved outflow identified in the Vul OB1 complex. The image shown has the following layout: (a) Continuum-subtracted H$_2$-K image, (b) J (blue), H (green) and H$_2$ (red) colour-composite image, (c) IRAC 3.6 $\mu$m (blue), 4.5 $\mu$m (green) and 8.0 $\mu$m (red) colour-composite image, (d) IRAC 4.5 $\mu$m (blue), 8.0 $\mu$m (green) and MIPS 24 $\mu$m (red) colour-composite image, (e) PACS 70 $\mu$m image with ATLASGAL 870 $\mu$m contours shown in grey and (f) SPIRE 250 $\mu$m (blue), 350 $\mu$m (green), 500 $\mu$m (red) colour composite image. The knots identified in the F15 catalog are shown by black regions. YSOs identified in the field are denoted by the following symbols: red square box - PBR, magenta square box - YSO identified using 24 $\mu$m and 70 $\mu$m colour, red, magenta and orange circles - Class 0/I, II and II YSOs identified from \citet{Billot_2010,Kim_2015}. The dashed red line connects all the knots linked to the outflow.}
  \label{fig:sample_multi}
\end{figure*}

Accurate identification and classification of the driving source candidates becomes crucial while estimating the physical properties such as length, luminosity, morphology and dynamical time of identified MHOs. As is well established, MHOs are driven by actively accreting YSOs that are distinguishable by their IR colours \citep[see][]{Varricatt_2010,Varricatt_2013,Froebrich_2016,Samal_2018}. Using {\it Spitzer} IRAC and MIPS band photometry and following the spectral index based classification criteria from \citet{Greene_1994}, the young stellar population in Vul OB1 and IRDC G53.2 have been classified by \citet{Billot_2010} and \citet{Kim_2015}, respectively. Along with these YSO catalogs, we create multiple colour-composite images, similar to those displayed in Figure \ref{fig:sample_multi}, for the identification of the driving source candidates for the identified MHOs. These include, (1) JHH$_2$, (2) IRAC 3.6 $\mu$m, 4.5 $\mu$m and 8.0 $\mu$m, (3) IRAC 4.5 $\mu$m, 8.0 $\mu$m and MIPS 24 $\mu$m, and (4) SPIRE 250 $\mu$m, 350 $\mu$m, 500 $\mu$m, colour-composite images. In addition, PACS 70 $\mu$m images overlaid with contours of ATLASGAL 870 $\mu$m emission are also utilized. Presence of cold dust emission traced at 870 $\mu$m near the jets/knots suggest that the driving sources are very likely embedded in dusty clumps or cores. YSOs identified from \citet{Billot_2010,Kim_2015} have been depicted by circles with Class 0/I YSOs in red, Class II YSOs in magenta, and Class III YSOs in orange. 

\begin{figure}
\centering
\includegraphics[width=0.45\textwidth]{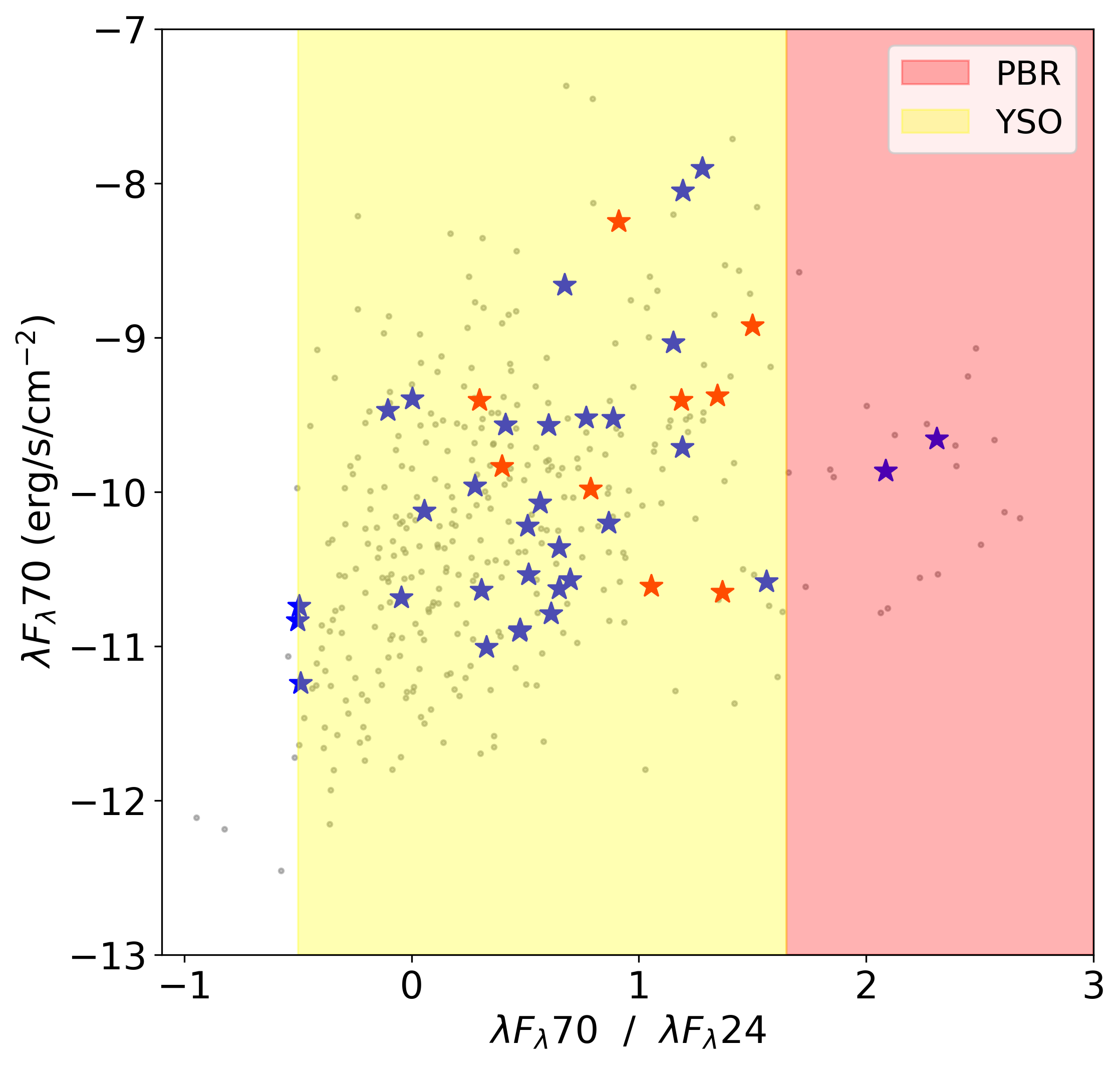}
\caption{Plot of 70 $\mu$m flux as a function of the ratio of 70 $\mu$m to 24 $\mu$m flux. HOPS protostars \citep{Stutz_2013,Manoj_2013} are shown as grey dots. The blue stars are the YSOs identified in this study, having detection in IRAC bands. Deeply embedded sources lacking detection in the IRAC bands are plotted as red stars. The yellow shaded region corresponds to log$\left(\frac{\lambda F_{\lambda} 70}{\lambda F_{\lambda} 24} \right) > -0.5$ which is used as the colour criterion for YSO identification. The red shaded regions represent PBRs, i.e., sources with log$\left(\frac{\lambda F_{\lambda} 70}{\lambda F_{\lambda} 24} \right) > 1.65$.}
\label{24_70}
\end{figure}

It should be noted that the deeply embedded protostars (e.g. Class~0 YSOs) are often faint or lack detection in the IRAC bands and hence may have been excluded from the YSO catalogues of \citet{Billot_2010,Kim_2015}. These sources, however, are bright at longer wavelengths and appear as bright sources in the MIPS 24 $\mu$m and PACS 70 $\mu$m bands (see Figure \ref{fig:sample_multi}). In order to identify these deeply embedded sources, we formulate a classification scheme based on their [24]-[70] colour. In Figure \ref{24_70}, we plot the protostars identified as part of the Herschel Orion Protostars Survey \citep[HOPS]{Stutz_2013,Manoj_2013}. As seen in the figure, majority of the HOPS protostars (98\%) are found to have colours, log$\left(\frac{\lambda F_{\lambda} 70}{\lambda F_{\lambda} 24} \right) > -0.5$. Using this colour cut, we identify 9 deeply embedded YSOs lacking detection in the {\it Spitzer} IRAC bands in Vul OB1 and IRDC G53.2. These are highlighted as red stars in the figure. YSOs with IRAC detection are also shown in the plot as blue stars. All the driving source candidates for outflows in this study, having both MIPS 24 $\mu$m and PACS 70 $\mu$m fluxes, satisfy the above-mentioned criterion. Protostars with extreme red colours satisfying log$\left(\frac{\lambda F_{\lambda} 70}{\lambda F_{\lambda} 24} \right) > 1.65$ are identified as PACS Bright Red sources or PBRs. These large infrared excess sources are believed to be early-Class 0 sources \citep{Stutz_2013}. In our study, we identify 2 PBRs based on their [24]-[70] colours. 

Using the generated colour-composite images and the catalog of YSOs and likely protostars, we identify the driving source candidate(s) based on a careful visual inspection of the location of YSOs with respect to the MHOs. For MHOs consisting of single knots, a driving source candidate is assigned based on any prominent extension of the H$_2$ emission features and proximity to a nearby YSO. For MHOs with multiple driving sources in vicinity, driving source candidates are assigned based on jet alignment. In such cases, preference has also been given to sources with a SiO detection in the catalogue of \citet{Csengeri_2016}. We acknowledge the possible subjectivity in the above classification process. More detailed studies based on other shock tracers can be used to confirm the proposed associations \citep[e.g.][]{Plunkett_2013,Zinchenko_2015}. Table \ref{Tab:Out_prop} shows the details of the outflows identified in this work and their physical parameters.

\begin{table*}
\setlength{\tabcolsep}{2pt}
\caption{Sample table showing details of MHOs identified in Vul OB1 and IRDC G53.2}
\label{Tab:Out_prop}
\begin{tabular}{@{}ccccccccccc@{}}
\toprule
\textbf{OF Name} & \textbf{MHO Name} & \textbf{\begin{tabular}[c]{@{}c@{}}R.A.$^{a}$\\ (J2000)\end{tabular}} & \textbf{\begin{tabular}[c]{@{}c@{}}Dec $^{a}$\\ (J2000)\end{tabular}} & \textbf{Cloud $^{b}$} & \textbf{Source Name $^{c}$} & \textbf{Morphology $^{d}$} & \textbf{\begin{tabular}[c]{@{}c@{}} Lobe Length\\ (pc) $^{e}$\end{tabular}} & \textbf{\begin{tabular}[c]{@{}c@{}}Lobe L$_{2.12}$\\ (L$_\odot$)$^{e}$\end{tabular}}
\\
\midrule
OF1	&	MHO 4218	&	19:41:44.0	&	23:14:12.0	&	Vul	&	G059.1854+00.1068	&	S-shaped, Bipolar	&	L: 0.18 R: 0.12	&	L: 0.001 R: 0.0009\\
OF2	&	MHO 4219	&	19:42:6.6	&	23:16:26.6	&	Vul	&	---	&	Linear, Unknown	&	---	&	---\\
OF3	&	MHO 4224	&	19:42:55.0	&	23:24:14.7	&	Vul	&	G059.4655-00.0457	&	Linear, Unipolar	&	1.8	&	0.0089\\
OF4	&	MHO 4216	&	19:41:1.2	&	22:3:9.9	&	Vul	&	G058.0750-00.3357	&	Single knot, Unipolar	&	0.8	&	0.002\\
OF5	&	MHO 4214	&	19:39:14.1	&	22:40:16.8	&	Vul	&	G058.4098+00.3279	&	Linear, Unipolar	&	0.35	&	0.004\\
OF6	&	MHO 2623	&	19:38:59.5	&	22:46:57.2	&	Vul	&	G058.4789+00.4318	&	Linear, Unipolar	&	1.19	&	0.0022\\
OF7	&	MHO 4210	&	19:38:37.3	&	22:41:16.5	&	Vul	&	---	&	Linear, Unknown	&	---	&	---\\
OF8	&	MHO 4208	&	19:38:33.3	&	22:43:5.0	&	Vul	&	J193833.26+224305.0	&	Linear, Unipolar	&	0.11	&	0.0052\\
OF9	&	MHO 4213	&	19:39:0.9	&	22:57:56.3	&	Vul	&	G058.6412+00.5168	&	Linear, Unipolar	&	0.07	&	0.0005\\
OF10	&	MHO 4209	&	19:38:36.5	&	23:4:43.8	&	Vul	&	---	&	Linear, Isolated knot	&	---	&	---\\
\midrule
\multicolumn{9}{p{\textwidth}}{(a) For outflows without any driving source candidate, the location of the outflows are listed as the mean of the coordinates of the farthest knots. For isolated knots, the location of the outflow corresponds to the coordinates knot itself.} \\ 
\multicolumn{9}{p{\textwidth}}{(b) Vul stands for Vulpecula OB association and IRDC is used to IRDC G53.2.} \\
\multicolumn{9}{p{\textwidth}}{(c) The source names correspond to the name of the driving source candidates in the PSC of various Galactic plane surveys. Sources names with prefix `G' indicate the coordinates of the point sources in GLIMPSE PSC \citep{GLIMPSE_PSC} in the Galactic coordinate system, `MG' corresponds to sources from MIPSGAL PSC \citep{Gutermuth_2015}. The prefix `HIGALPB' and `HIGALPR' indicates point sources from the Hi-GAL PSC \citep{Molinari_2016} in PACS 70 $\mu$m and PACS 160 $\mu$m bands, respectively. Source with prefix `J' correspond to WISE PSC \citep{WISE_PSC} and indicate the coordinates of the sources in equatorial system (J2000). The location of the outflows as listed in the table corresponds to the coordinates of the driving source candidate, if identified.} \\
\multicolumn{9}{p{\textwidth}}{(d) This column lists the geometry and the nature of the outflow, i.e. unipolar, bipolar or multi-polar (more than one outflow).} \\
\multicolumn{9}{p{\textwidth}}{(e) In case of bipolar outflows, 'L' is used to indicate properties of the left lobe and 'R' represents properties of right lobe.} \\ \midrule
\multicolumn{9}{r}{\bf Full table available in electronic version.} \\
\bottomrule
\end{tabular}
\end{table*}

Based on the above approach, driving source candidates are identified for 79 (62\%) outflows. Similar statistics is seen in other Galactic star-forming regions, where only 50-60\% of the outflows were found to have associated driving sources \citep[see][and references therein]{Samal_2018}. Of the 79 outflows with identified driving sources, we find that 27 MHOs are bipolar in nature and 52 are unipolar. The lower number of bipolar outflows identified could be attributed to high extinction which limits the detection of the red-shifted lobes in some cases. Additionally, other factors such as anisotropic ambient medium \citep[e.g.][]{Lopez_2013} could lead to a higher unipolar outflow statistics compared to the bipolar counterparts as seen in this study.

\subsection{Extinction and distance to identified MHOs}\label{PNICER}

\begin{figure*}
  \begin{subfigure}[t]{0.60\textwidth}
  \includegraphics[width=\textwidth]{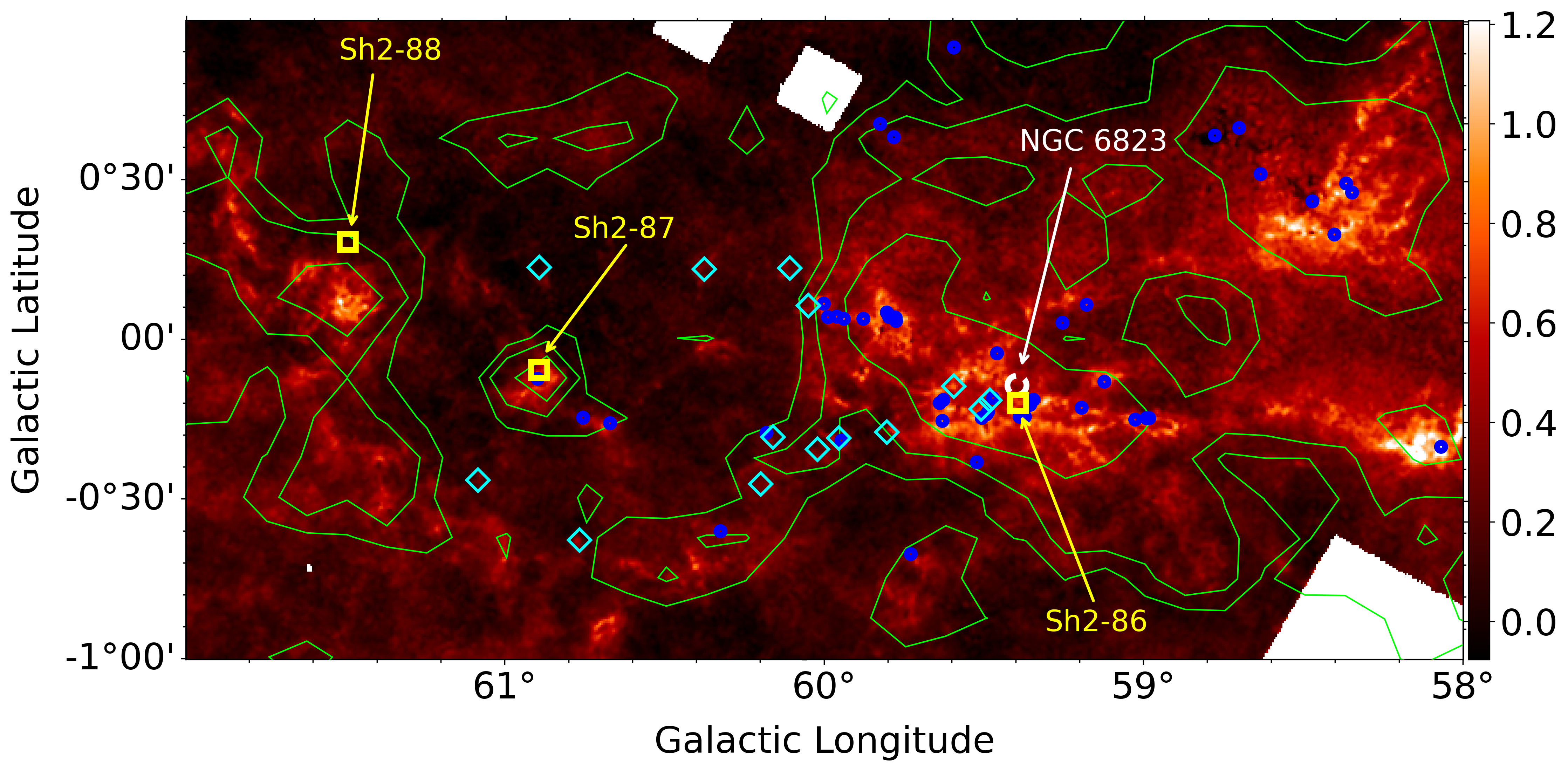}
    \caption{}
    \label{fig:Vulp_ext_CO}
  \end{subfigure}%
  ~
  \begin{subfigure}[t]{0.355\textwidth}
    \includegraphics[width=\textwidth]{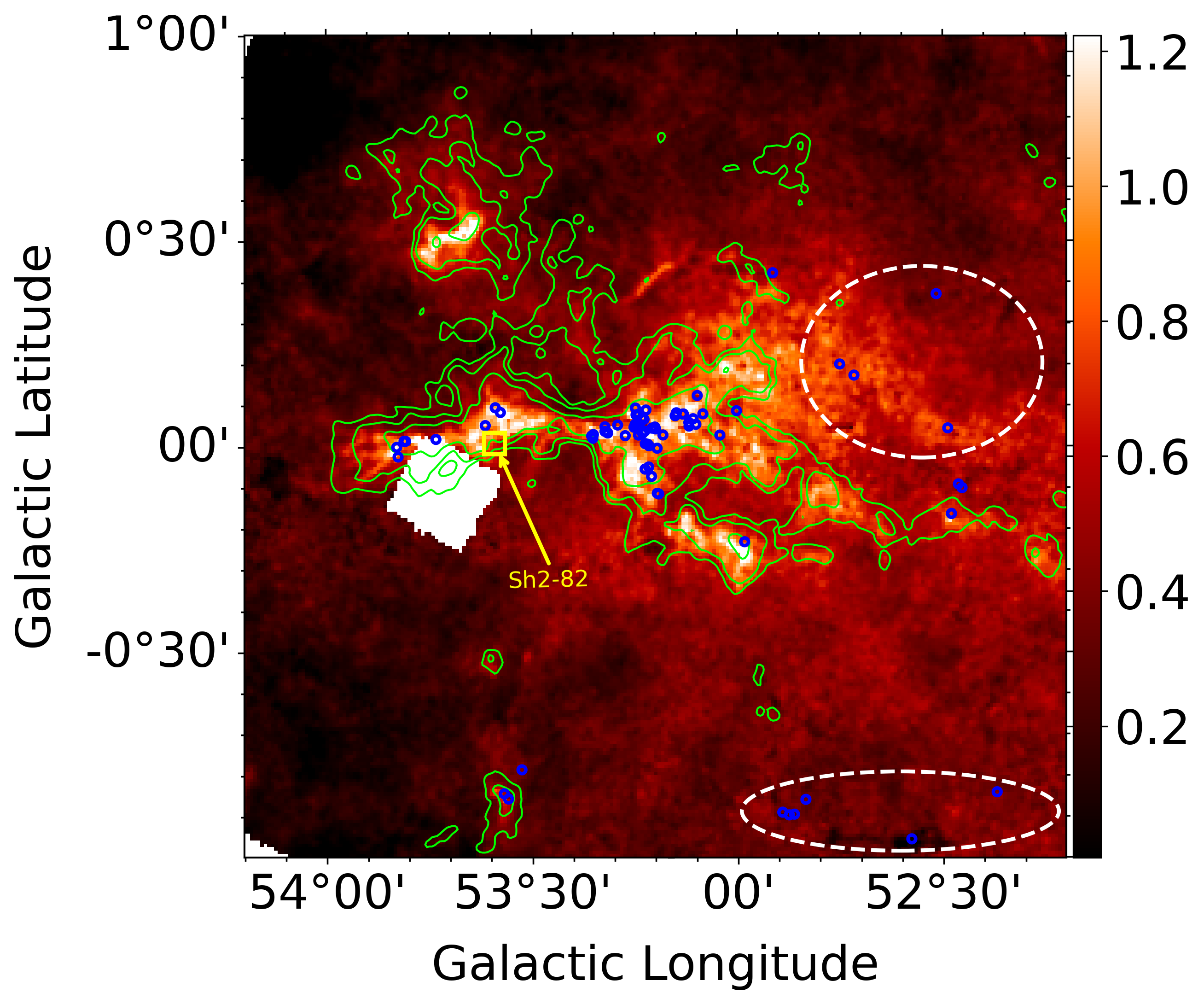}
    \caption{}
    \label{fig:IRDC_ext_CO}
  \end{subfigure}
  \caption{Extinction map for (a) Vul OB1 and (b) IRDC G53.2 generated using PNICER with a resolution of 1\arcmin. The blue dots show the locations of the MHOs identified using the continuum-subtracted images and the green contours represent the 3$\sigma$, 5$\sigma$ and 10$\sigma$ boundaries from the CO velocity-integrated map from \citet{Dame_2001} and \citet{Jackson_2006}. The location of various pillars in Vul OB identified by \citet{Billot_2010} are shown by cyan diamonds. HII regions from \citet{Sharpless_1959} are depicted by yellow box. The white dashed regions in (b) show the MHOs outside the boundary of IRDC G53.2. The empty square corresponds to regions lacking data in the GPS survey.}
  \label{fig:ext_map}
\end{figure*}

Extinction and distance information to individual MHOs is required in order to estimate the physical parameters. Extinction maps for the two clouds are generated using the PNICER algorithm developed by \citet{Meingast_2017} \footnote{\url{http://smeingast.github.io/PNICER/}}. Photometric data from the UKIDSS GPS point source catalog \citep{Lucas_2008} for the target area referred to as the science field here) and an extinction-free control field are provided as input to PNICER. The control field is identified as regions devoid of dust emission in the AKARI-FIR colour-composite images using Aladin sky atlas \citep{Bonnarel_2000}. PNICER calculates the colour excess for point sources in the science field using the colours of stars inside the control field and returns a discrete extinction map at K-band. Subsequently, the maps are smoothed to a resolution of 30\arcsec with pixel scale of 15\arcsec (i.e. 0.12~pc and 0.17~pc corresponding to a distance of 1.7 and 2.3~kpc, respectively). The generated extinction maps are presented in Figure \ref{fig:Vulp_ext_CO} and \ref{fig:IRDC_ext_CO} for Vul OB1 and IRDC G53.2, respectively. Since the pixel scale is much larger than the typical size of H$_2$ knots, we use the value of the pixel corresponding to the knot as our estimate of extinction. Few regions of high extinction and hence with smaller stellar densities compared to our control field appear as dark patches in the extinction map. For such cases, we use a lower resolution 1\arcmin and pixel scale of 30\arcsec extinction map. We note, due to beam dilution, the extinction values in dense compact regions are likely to be underestimated.

CO velocity-integrated contours, generated using CO data cubes from \citet{Dame_2001,Jackson_2006}, are overlaid in Figure \ref{fig:Vulp_ext_CO} and \ref{fig:IRDC_ext_CO}. The contours represent emissions between cloud velocity ranges of $v=20-40$ kms$^{-1}$ and $v=15-30$ kms$^{-1}$ taken from \citet{Billot_2010} and \citet{Kim_2015} for the Vul OB1 and IRDC G53.2 complexes, respectively. The outermost contours, corresponding to the 3$\sigma$ levels, define approximate boundaries for the complexes. MHOs lying within and close to the 3$\sigma$ emissions are considered to be associated with the parent cloud complex. In the absence of distance information for individual outflows, we assume a common distance for all outflows within the same cloud complex. As seen in Figure \ref{fig:ext_map}, the majority of the detected MHOs lie within the cloud boundaries defined by the 3$\sigma$ levels, with some exceptions in the IRDC G53.2. While the MHOs lying outside the cloud boundaries from CO maps may still be part of the cloud, they have been excluded from further analysis. 

It should be kept in mind that physical association of individual MHOs is difficult to confirm from this plane-of-sky correlation, as it is not possible to distinguish MHOs driven by foreground/background YSOs. Based on the CO data, we find no other major cloud components along the line-of-sight of both the complexes. Hence, we assume that the contamination from other Galactic MHOs would be minimal in our identified sample.

\begin{figure*}
  \begin{subfigure}{0.33\textwidth}
    \centering
    \includegraphics[width=\textwidth]{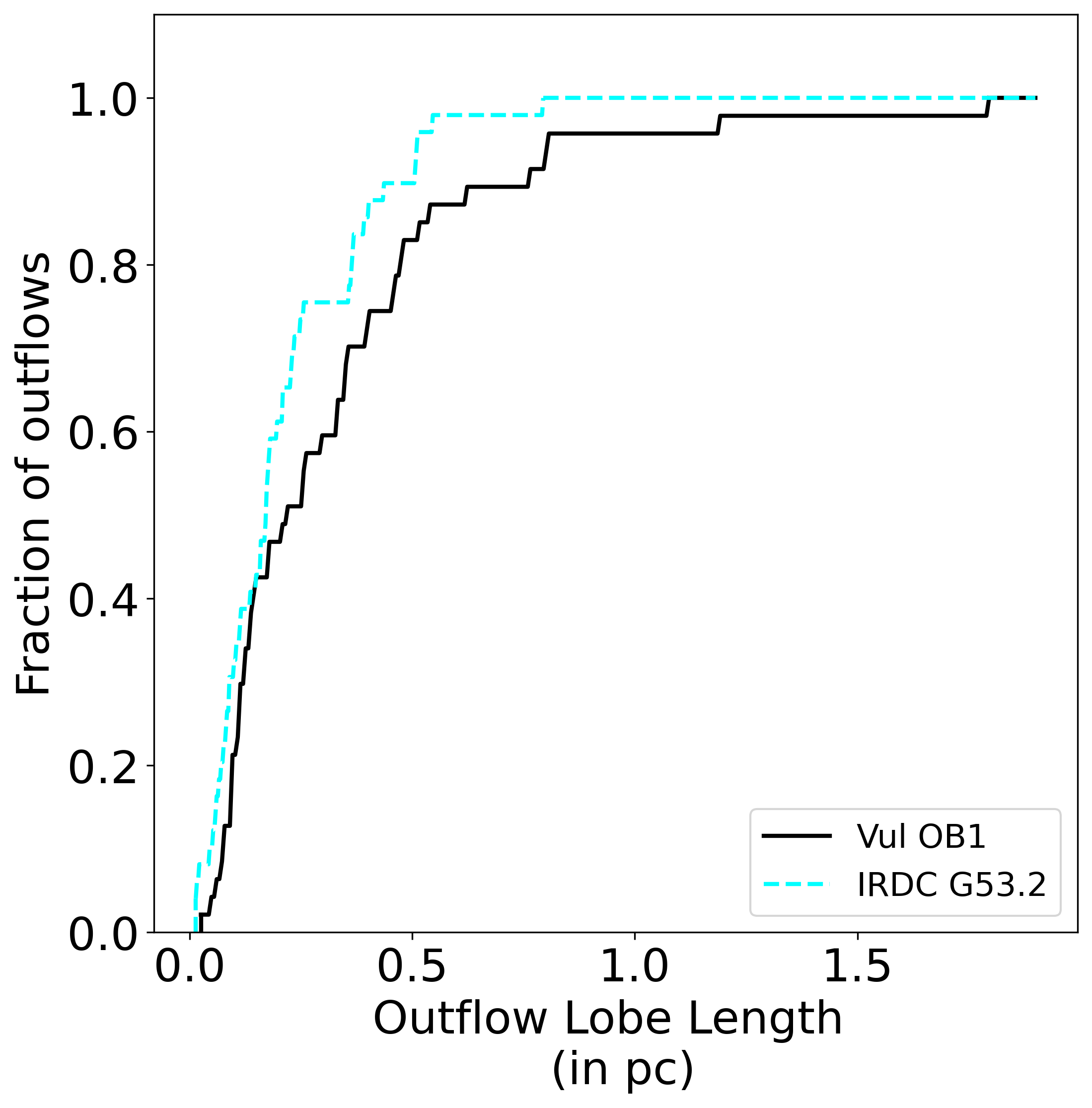}
    \caption{}
    \label{fig:out_len_dist}
  \end{subfigure}%
  \begin{subfigure}{0.33\textwidth}
    \centering
    \includegraphics[width=\textwidth]{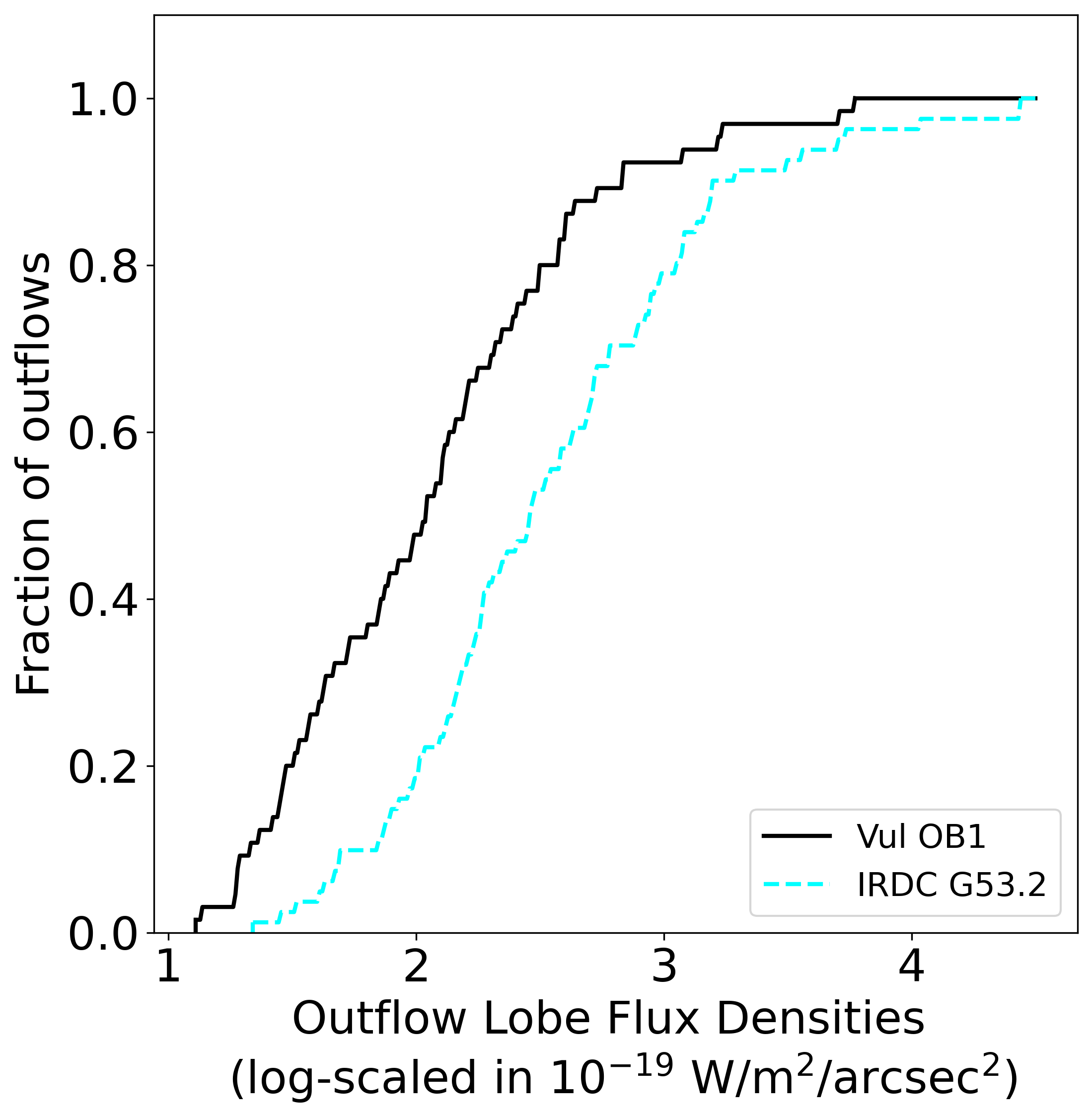}
    \caption{}
    \label{fig:out_flux_dist}
  \end{subfigure}%
  \begin{subfigure}{0.33\textwidth}
    \centering
    \includegraphics[width=\textwidth]{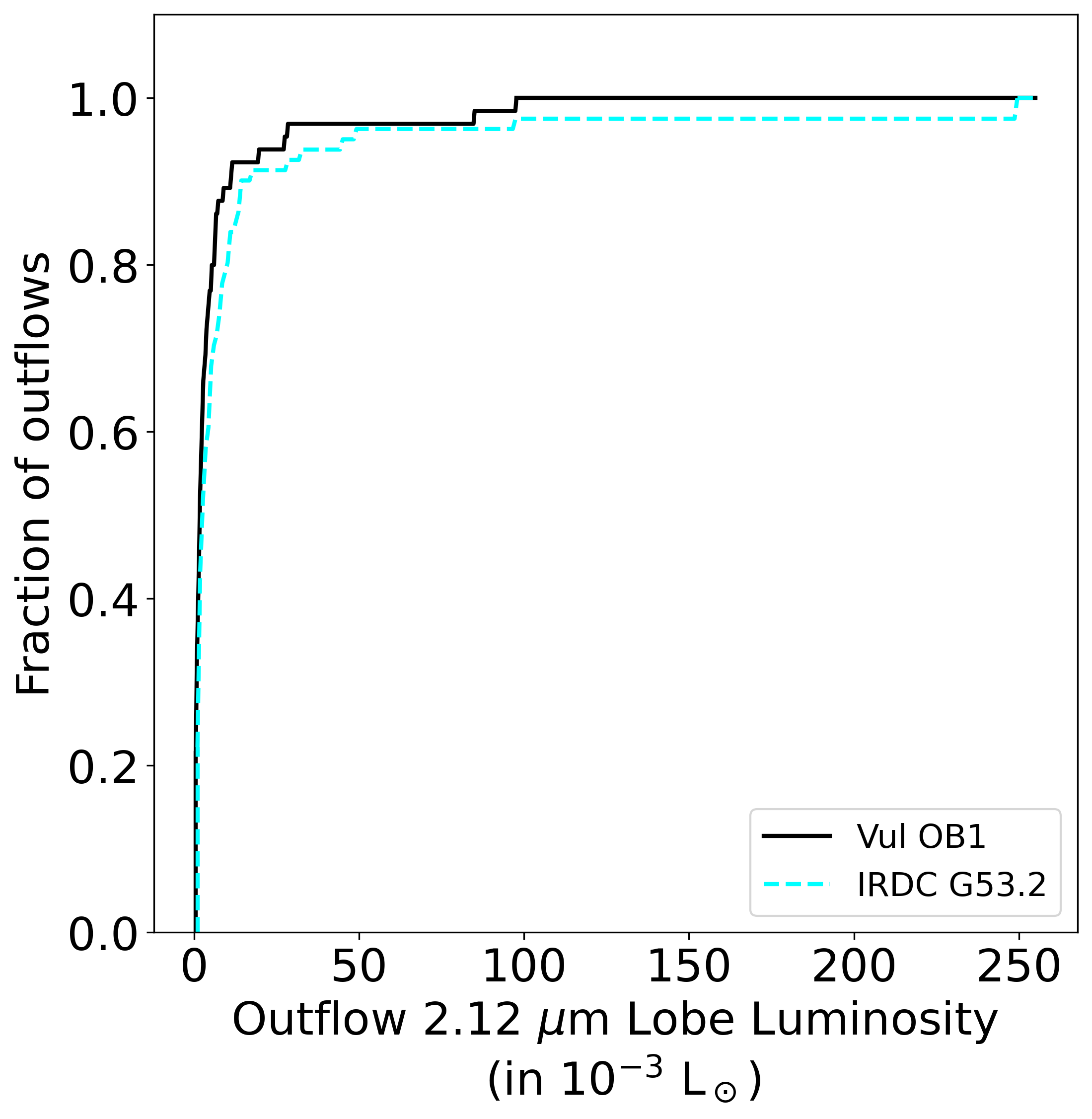}
    \caption{}
    \label{fig:out_lum_dist}
  \end{subfigure}
  \caption{Cumulative distribution function of (a) outflow lobe length, (b) outflow lobe H$_2$ 1-0S(1) flux density and (c) H$_2$ 1-0S(1) line-luminosity of outflow lobes in Vul OB1 and IRDC G53.2.}
  \label{fig:out_len_flux_lum}
\end{figure*}

\subsection{Outflow lobe length and H$_2$ 1-0S(1) line-luminosity distribution}\label{sec:out_len_lum}

Following the approach of \citet{Makin_2018}, we consider individual outflow lobes irrespective of bipolar or unipolar outflow identification for estimating the lobe length. It should be noted that these estimates for outflow lobe length have not been corrected for inclination and thus should be treated as lower limits. Figure \ref{fig:out_len_dist} shows the number distribution of the lobe length estimates. The distribution yields a consistently larger median lobe length of 0.22~pc for Vul OB1 as compared to 0.17~pc for IRDC G53.2. In comparison, they are similar in angular units ($\sim$20 arcsec), thus indicating the effect of distance on the physical lengths. The observed lobe lengths are consistent with the range of $\sim$ 0.2 - 0.4~pc seen in other complexes like Aquila, Orion A, Cassiopeia and Auriga, Serpens and Aquila, Cygnus-X and M17 \citep{Zhang_2013,Stanke_2002,Ioannidis_2012a,Froebrich_2016,Makin_2018,Samal_2018}. 

From observations of several clouds, it is seen that long, parsec-scale outflows are few \citep{Davis_2009,Ioannidis_2012a,Stanke_2002,Makin_2018}. Considering both complexes, around 10\% of the outflow lobes with known driving source candidates are estimated to be longer than 0.5~pc and 7 outflows have a total length greater than 1~pc. However, the actual number of parsec-scaled outflows may be higher if corrected for the effect of outflow inclination. Furthermore, the H$_2$ 1-0S(1) line traces only the hot and dense region of the outflows and has a fast cooling rate. Thus, some of the evolved and faint emission features might lack detection, especially at large distances. Sensitive optical observations in the H$_\alpha$ and [SII] lines are required to trace distant outflows in low density regions of the cloud. 

The flux densities of each H$_2$ emission knot are retrieved from the F15 catalog. The retrieved flux densities are then de-reddened using the following expression
\begin{equation}
  F_{\rm dered}=F_0 \times 10^{\frac{A_{2.12\mu m}}{2.5}}\\
\end{equation}  
where $F_{\rm dered}$ is the de-reddened knot flux, $F_0$ is the knot flux at 2.12 $\mu$m and $A_{2.12\mu m}$ is the extinction value at 2.12~$\mu$m calculated using the A$_K$ value from the extinction map \citep[$A_{2.12}$=~1.06~$\times$~$A_K$ from][]{Cardelli_1989}. Subsequently, the H$_2$ 1-0S(1) line flux density and luminosity of the identified lobes are calculated. For a typical jet/outflow temperature of 2200~K, the H$_2$ 1-0S(1) line luminosity accounts for roughly one-twelfth the total H$_2$ luminosity emitted from shock-excited H$_2$ \citep{Garatti_2006}. 

Figure \ref{fig:out_flux_dist} and \ref{fig:out_lum_dist} show the distribution of the outflow lobe flux densities and the lobe H$_2$ line-luminosity, respectively. The outflow lobe flux densities vary between $\rm 1.2\times 10^{-18}$ to $\rm 2.7\times 10^{-15}$W/m$^2$/arcsec$^2$ with median value of 1.1$\times 10^{-17}$W/m$^2$/arcsec$^2$ and 2.9$\times 10^{-17}$W/m$^2$/arcsec$^2$ for Vul OB1 and IRDC G53.2 respectively.

It should be noted that, the flux density of the faintest H$_2$ knot identified in this study (5.9$\times$10$^{-19}$ W/m$^2$/arcsec$^2$) is comparable to the detection limit of H$_2$ features in surveys of nearby clouds \citep[ $\rm 1~-~7\times10^{-19}W/m^2/arcsec^2$ in][]{Stanke_2002,Zhang_2015,Davis_2009,Zhang_2013}. However, since sources in our study are located significantly farther away, the faintest outflow identified in this study is nearly 17 times more luminous than the faintest outflow in Orion A \citep{Stanke_2002} and 233 times more brighter compared to the faintest outflow in Aquila molecular cloud \citep{Zhang_2015}. These differences suggest a possible bias towards more luminous outflows from intermediate- and high-mass stars in our study.

\subsection{Distribution of YSO mass and luminosity from SED models}

\citet{Qiang_2018} show that the mechanical force of outflows from protostars is well correlated with the bolometric luminosity of the central source and the correlation holds good over the entire mass regime. The bolometric luminosity of protostars is mainly dominated by energy release associated with accretion of matter. However, for late stage YSOs, the photometric luminosity dominates the accretion luminosity. In order to obtain reliable estimates of photometric luminosity of the driving source candidates, we use the SED-fitting tool\footnote{\url{https://sedfitter.readthedocs.io/en/stable/}} developed by \citet[][ hereafter R17]{Robitaille_2017}. The SED-fitter uses $\chi^2$-minimization to fit the observed SEDs of YSOs to a large grid of YSO model SEDs spanning a large parameter space for physical properties like dust distribution, radius and temperature of central star, etc. The R17 model addresses some of the caveats of the SED models from \citet[][ hereafter R06]{Robitaille_2006} by eliminating highly model-dependent parameters which do not directly affect the SEDs, such as mass, luminosity and envelope infall rates. The R17 models also provide a wider and uniformly distributed parameters compared to the R06 models. The R17 models comprise of 18 different models of varying complexities with different combinations of YSO parameters, offering modularity to the users in choosing the model that best represent the source parameters. These models are named using characters representing presence of different components in the model, for instance, the model `spubhmi' corresponds to a source with a central star (s) with a passive disk (p), ulrich envelope (u), bipolar cavities (b), inner hole (h), ambient medium (m) and interstellar dust (i) (refer to R17 for a detailed description of all 18 models).

Photometric data, covering the wavelength range of 1.2-70 $\mu$m, taken from UKIDSS-GPS, GLIMPSE, MIPSGAL, and Hi-GAL catalogs \citep{GLIMPSE_PSC,Gutermuth_2015,Lucas_2008,Molinari_2016} are provided as inputs to the SED-fitting tool. Based on distance estimates from literature \citep{Billot_2010,Kim_2015} and keeping in mind the uncertainties associated with these values, we use a conservative distance range of 2.0 to 2.5 kpc for Vul OB1 and 1.5 to 2.0 kpc for IRDC G53.2 for the SED modelling. Visual extinction input is allowed to vary from 2 to 50 mag. Taking 1 mag per kpc increase in A$_v$ \citep{Stahler_2004} sets the lower limit of the foreground extinction in the direction of both the clouds, while the typical maximum extinction seen towards extended green objects (EGOs) or ‘green fuzzies’ \citep[EGOs]{Cyganowski_2008} and Ultracompact HII regions \citep[UCHII,][]{Hanson_2002,Garatti_2015} is taken as the upper limit.

\begin{figure}
\centering
  \begin{subfigure}{0.45\textwidth}
    \includegraphics[width=\textwidth]{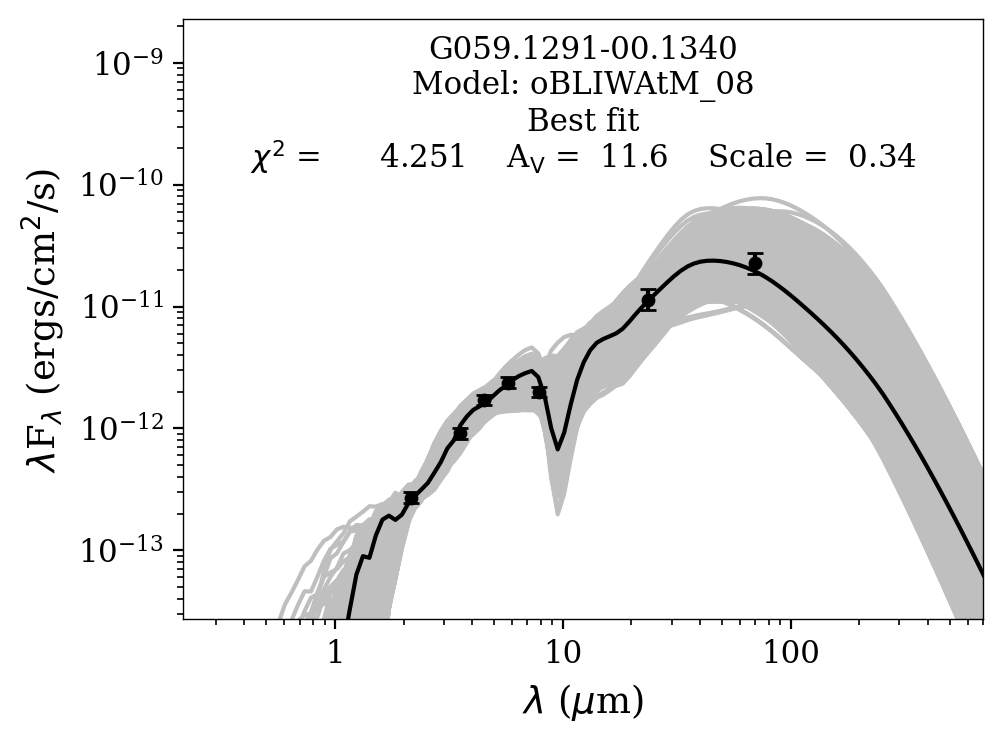}
    \caption{SED fitting for a Class 0/I YSO.}
    \label{fig:SED_1}
  \end{subfigure}%
  \\
  \begin{subfigure}{0.45\textwidth}
    \includegraphics[width=\textwidth]{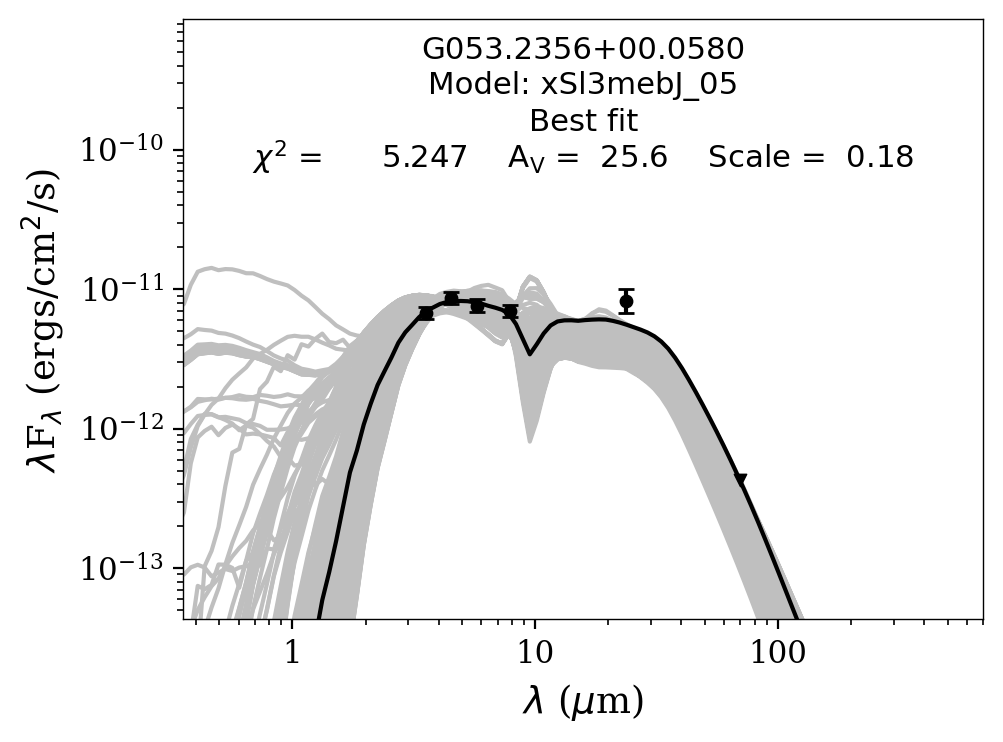}
    \caption{SED fitting for a Class II YSO.}
    \label{fig:SED_2}
  \end{subfigure}
  \caption{Sample output from SED fitting using R17 models showing the good-fit models that satisfy the $\chi^2-\chi^2_{min}\leq9N_{data}$ criteria. The black dots are the observed photometric data points given as input. The black SED corresponds to the best-fit model, and the gray SEDs are for the remaining models that are consistent with the input fluxes. It should be noted that $\chi^2_{min}$ represents the least $\chi^2$ for the source across all 18 SED models.}
  \label{fig:SED}
\end{figure}

\begin{figure}
\centering
  \includegraphics[width=0.45\textwidth]{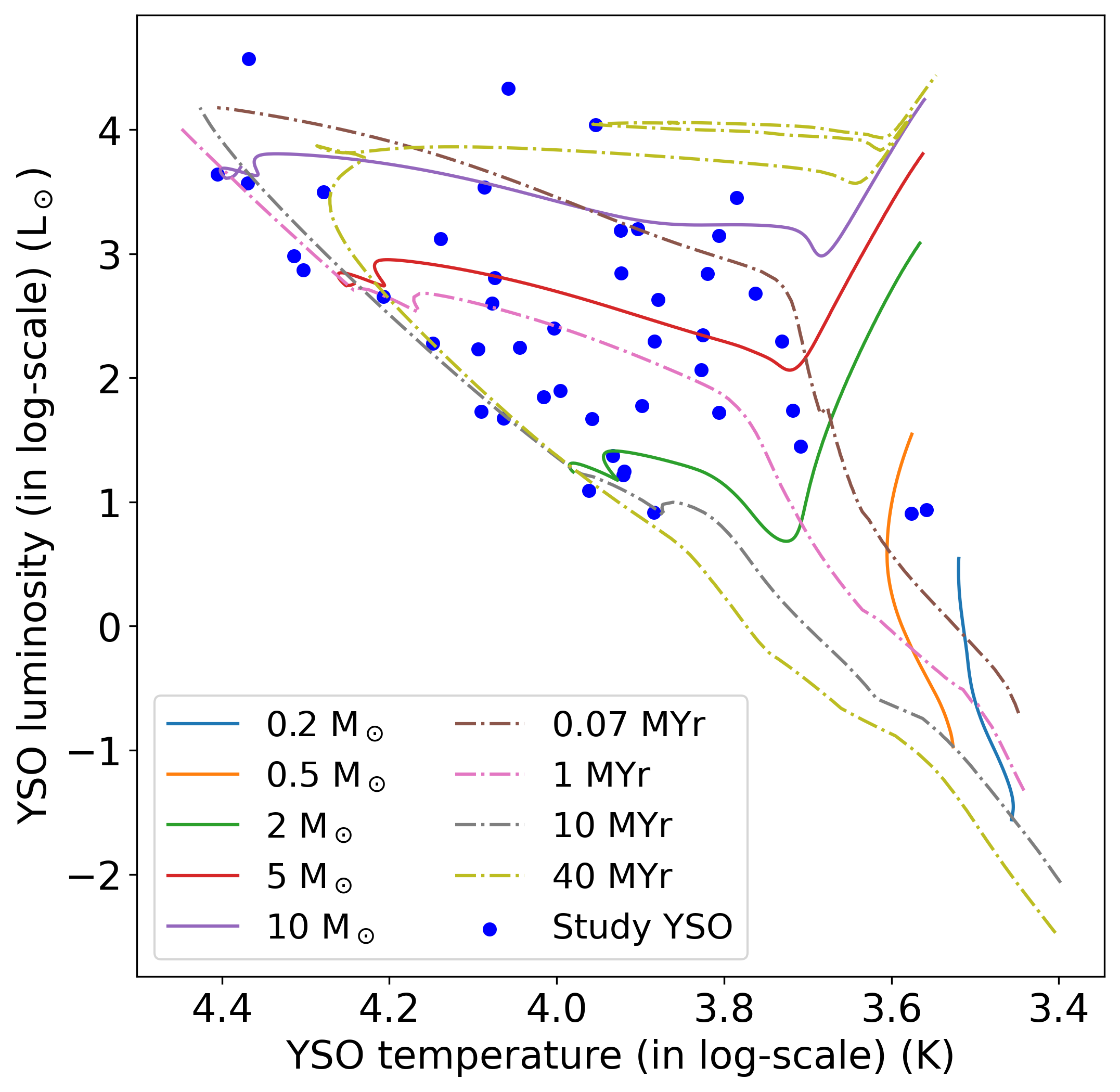}
  \caption{HR diagram of driving source candidates for MHOs identified in this study. Solid lines represent the evolutionary track of sources with the same mass and dashed lines represent the isochrone tracks. The blue dots represent the driving source candidates identified in this study.}
  \label{fig:YSO_HR}
\end{figure}

\begin{figure}
\centering
  \begin{subfigure}{0.45\textwidth}
    \includegraphics[width=\textwidth]{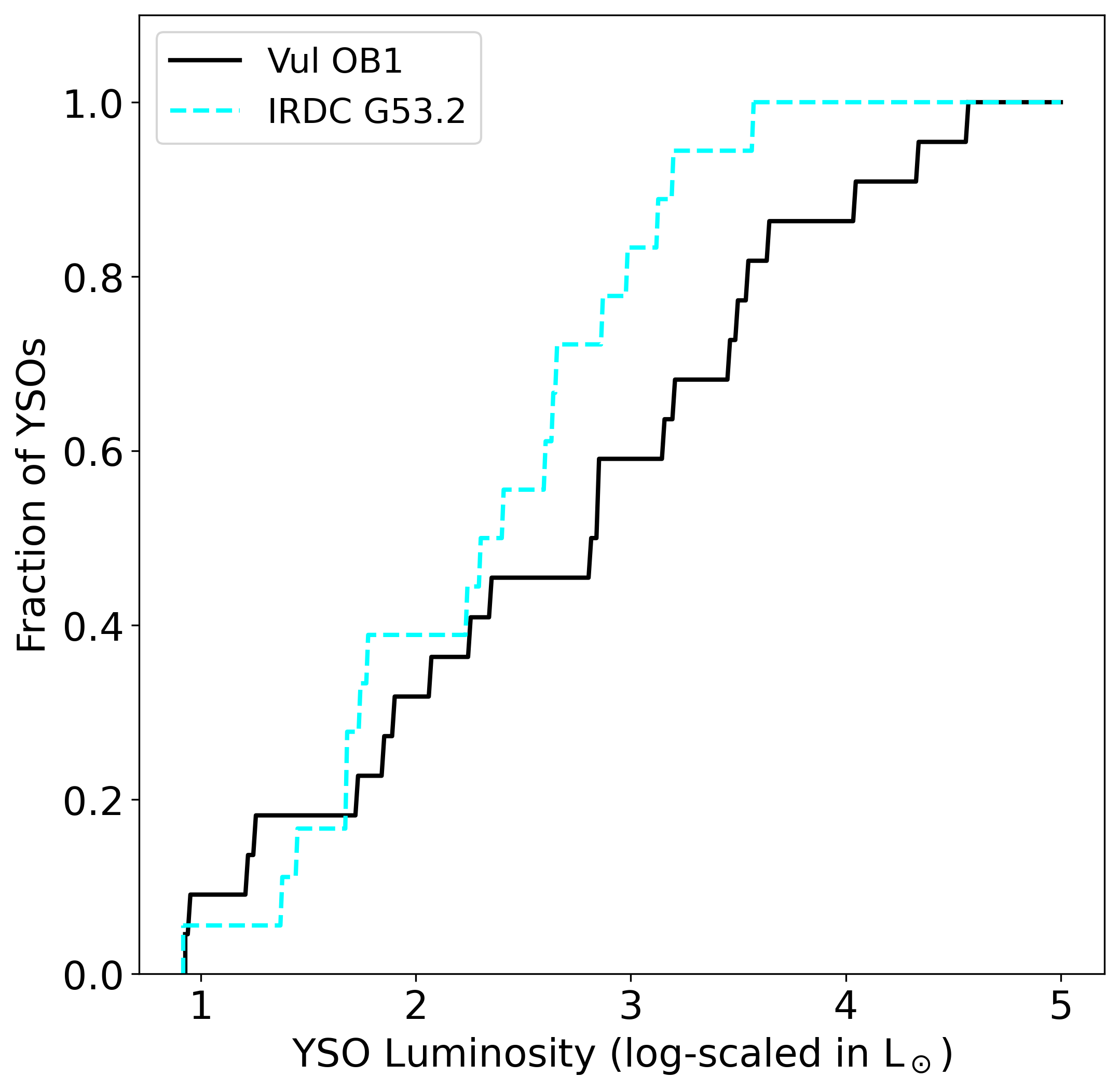}
    \caption{}
    \label{fig:YSO_Lum}
  \end{subfigure}%
  \\
  \begin{subfigure}{0.45\textwidth}
    \includegraphics[width=\textwidth]{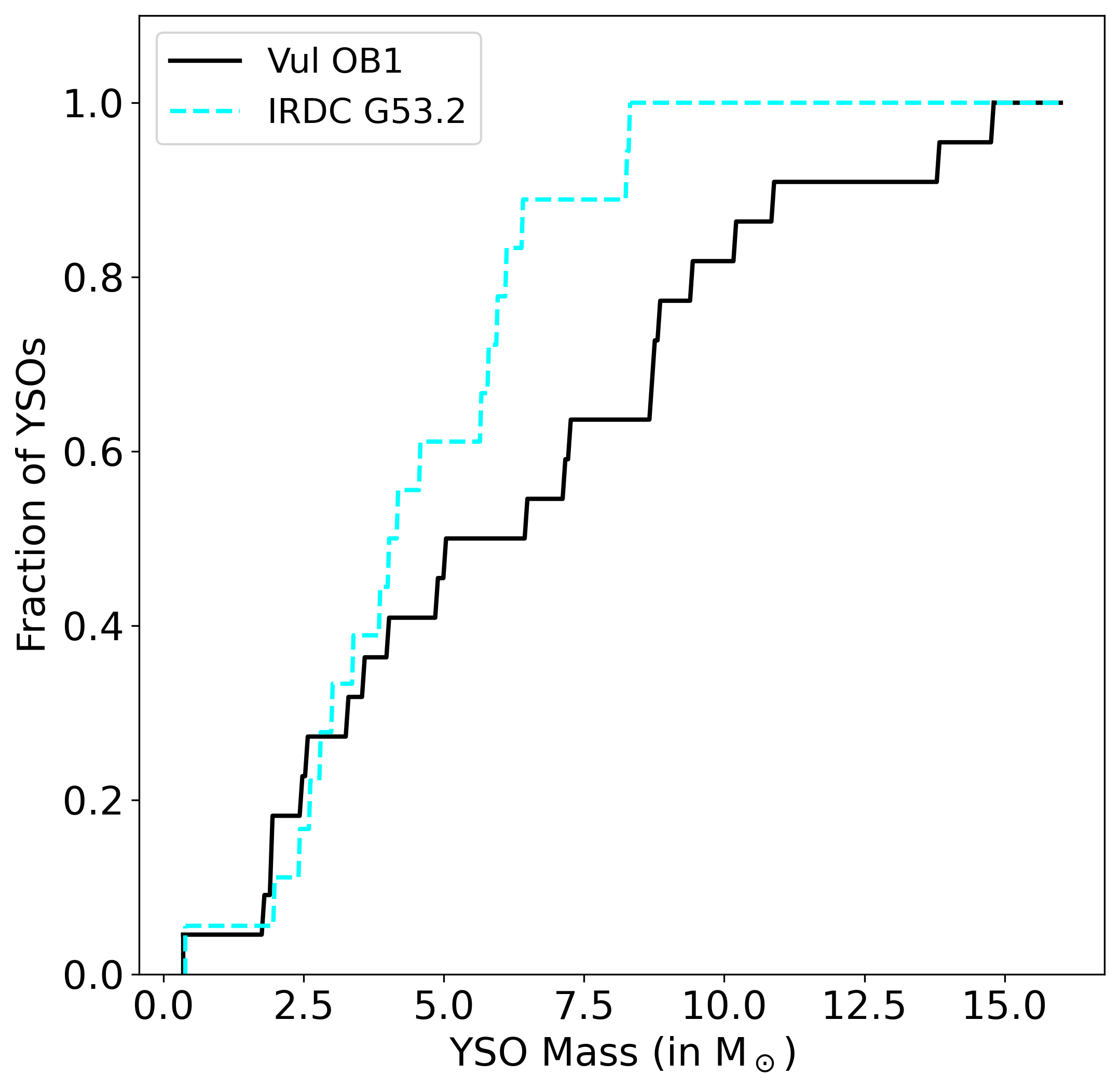}
    \caption{}
    \label{fig:YSO_mass}
  \end{subfigure}
  \caption{Cumulative distribution function of (a) luminosity and (b) mass of driving source candidates in Vul OB1 and IRDC G53.2.}
  \label{fig:YSO_mass_lum}
\end{figure}

Longer wavelength (24 and 70 $\mu$m) flux densities are important for constraining the fitted SED models \citep{Dunham_2006}. For undetected and saturated sources in these bands, we use the faintest and brightest flux densities from the point source catalog (see Section \ref{Sec:Data}) and use them as upper and lower limits, respectively. Similar to \citet{Samal_2018}, we use a conservative 10\% uncertainty on the flux density values of UKIDSS and GLIMPSE surveys and 20\% uncertainty on MIPSGAL and Hi-GAL flux density measurements. In order to avoid any bias,  we follow the approach described in the R17 paper. For every YSO, SED-fitting is carried for all 18 models and the minimum $\chi^2$ value ($\chi^2_{\rm min}$) for the best-fit model across all 18 models is determined. A relative score for each model is generated based on the likelihood of the model (P(D|M) = fraction of good-fit SEDs identified relative to the total number of SEDs in the model) to provide a fit to the input flux densities with a broadened threshold of {$\chi^2-\chi^2_{\rm min}\leq9$N$_{\rm data}$}. Here, N$_{\rm data}$ is the number of photometric data points used. Sample SEDs and the fitted models for a Class 0/I and a Class II YSO are shown in Figure \ref{fig:SED}. The physical parameters of each source are determined by using the model with the highest score. Exponential weights (e$^{-\frac{\chi^2}{2}}$) are assigned to the retrieved parameters of the chosen model, such that model fits with lower $\chi^2$ values are given a higher weight. The weighted mean and the corresponding error (standard deviation) of the model-derived physical parameters are listed in Table. \ref{Tab:YSO_prop}. Since the R17 models returns only the temperature and radius of the YSO as parameters of the central source , we used the Stephen-Boltzmann law to estimate the photometric luminosity of the YSOs. 

For estimating the mass of the sources, we use the Parsec Isochrone Datasets \citep{Bressan_2012} of solar metallicity and mass corresponding to the closest evolutionary track in the HR diagram is used as the mass of the YSO. Figure \ref{fig:YSO_HR} shows the HR diagram for the identified driving source candidates. We note that stellar parameters of YSOs are better constrained by SED models when optical measurements are available, which is not the case for the YSOs in our sample. Hence, these estimated values should be considered as indicative only. Additional high-sensitivity optical and infrared photometric observations are needed to accurately measure the stellar properties of the sources. Moreover, young pre-main-sequence (PMS) stars, in particular those associated with jets/outflows, are highly variable. Variability can induce enormous changes in their position in the HR diagram. In the present work, we reject those sources that lie significantly outwards of the boundaries defined by the HR diagram. In the case of high to intermediate stars, since PMS evolutionary track of a given mass star is nearly horizontal for different ages, thus, effect of uncertainty in age has less impact on the mass estimation.

The cumulative distribution functions of the mass and luminosity of the 45 selected driving source candidates are shown in Figure \ref{fig:YSO_mass_lum}. The estimated mass and luminosity of the driving source candidates range between $\sim$ 0.1-15 M$_\odot$ and $\sim$ 2-37226 L$_\odot$, respectively. The median values of the derived mass and luminosity are $\sim$ 4.4 M$_\odot$ and $\sim$ 221 L$_\odot$, respectively. While 10 sources are candidate massive YSOs (> 8 M$_\odot$), the median value suggests that the majority of the outflows identified in this work are likely driven by intermediate mass stars. From SED-fitting, we find that majority of the YSOs have an inclination angle of 40-50 degrees relative to the plane of the sky. Assuming a typical inclination angle of 45 degrees, our estimate of outflow lobe lengths could be under-estimated by a factor of only $\sim$1.4.

\begin{table*}
\caption{Sample table showing details of physical properties of driving source candidates of MHOs identified in Vul OB1 and IRDC G53.2}
\label{Tab:YSO_prop}
\begin{tabular}{@{}cccccc@{}}
\hline
\textbf{OF Name} & \textbf{MHO Name} & \textbf{Source Name} & \textbf{YSO temperature (K)} & \textbf{YSO luminosity (L$_\odot$)} & \textbf{YSO mass (M$_\odot$)}\\
\hline
OF1	&	MHO 4218	&	G059.1854+00.1068	&	11079$\pm$3926	&	176$\pm$102	&	3.6\\
OF2	&	MHO 4219	&	---	&	---	&	---	&	---\\
OF3	&	MHO 4224	&	G059.4655-00.0457	&	12208$\pm$5861	&	3441$\pm$2757	&	8.8\\
OF4	&	MHO 4216	&	G058.0750-00.3357	&	6607$\pm$4185	&	693$\pm$596	&	7.2\\
OF5	&	MHO 4214	&	G058.4098+00.3279	&	7997$\pm$6690	&	1587$\pm$1097	&	8.7\\
OF6	&	MHO 2623	&	G058.4789+00.4318	&	6685$\pm$291	&	222$\pm$31	&	5.0\\
OF7	&	MHO 4210	&	---	&	---	&	---	&	---\\
OF8	&	MHO 4208	&	J193833.26+224305.0	&	---	&	---	&	---\\
OF9	&	MHO 4213	&	G058.6412+00.5168	&	---	&	---	&	---\\
OF10	&	MHO 4209	&	---	&	---	&	---	&	---\\
\hline
\multicolumn{6}{r}{\textbf{Full table available in electronic version.}}\\
\hline
\end{tabular}
\end{table*}

\section{Discussion}\label{Sec:Discussion}

\subsection{Dominant evolutionary class for jet-bearing YSOs}
In both complexes, the identified YSOs have been classified using various photometric criteria in literature \citep{Billot_2010,Kim_2015}. However, for uniformity, we classify the identified driving source candidate YSOs based on the infrared spectral index, $\alpha_{\rm IR} = d$log$(\lambda F_\lambda)/ d$log$(\lambda)$ \citep[][]{Lada_1987}. $\alpha$ is estimated as the slope of the SED of the YSOs using the IRAC bands (3.6-8.0 $\mu$m). Following \citet{Greene_1994} and \citet{Billot_2010}, we classify the YSOs, as Class 0/I ($\alpha_{\rm IRAC} > -0.3$), Class II ( $-1.6 <\alpha_{\rm IRAC} < -0.3$) and Class III YSOs ( $-2.56 <\alpha_{\rm IRAC} < -1.6$).

\begin{figure}
  \centering
  \includegraphics[width=0.45\textwidth]{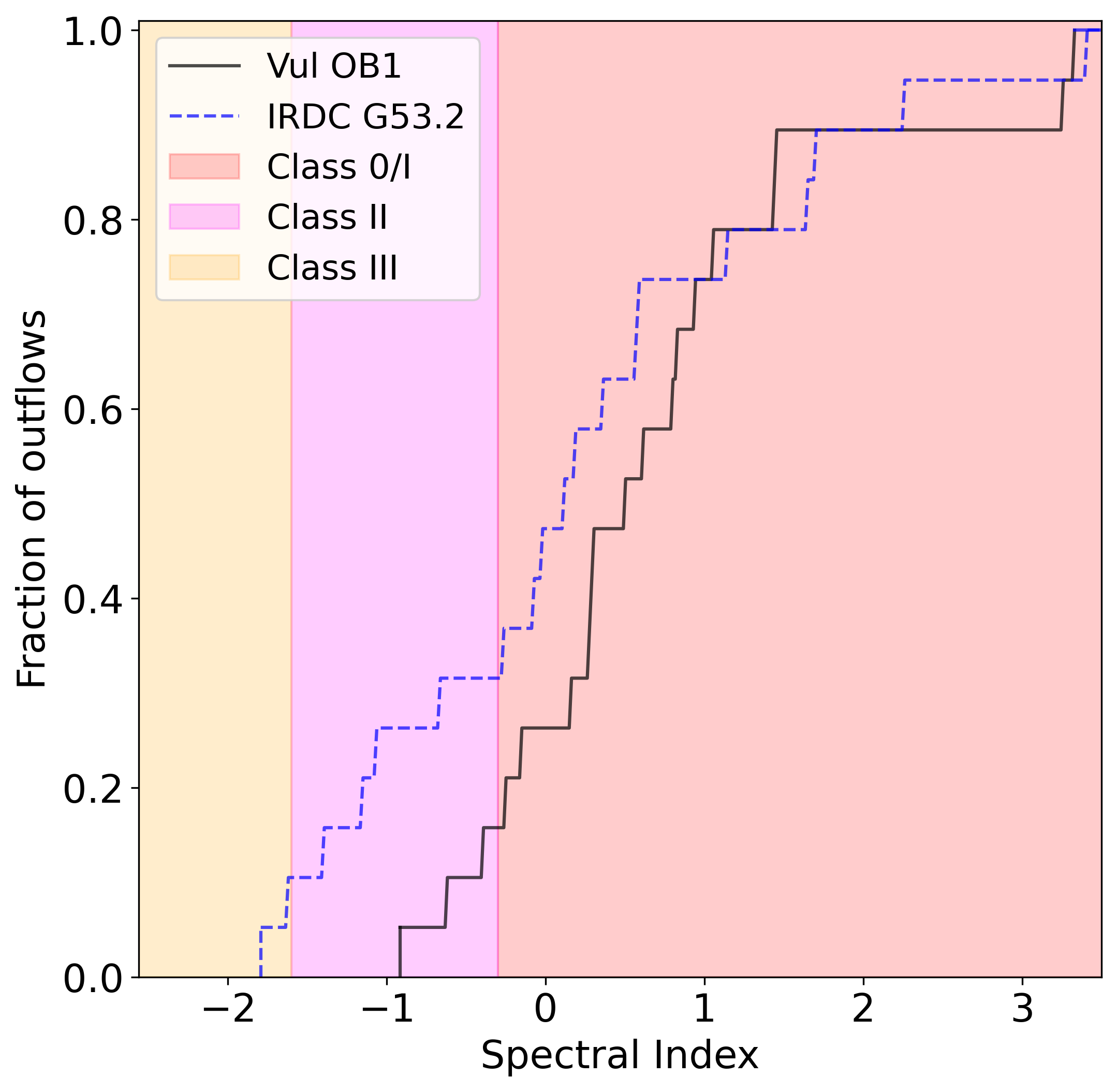}
  \caption{Cumulative distribution function of spectral index ($\alpha$) of driving source candidates for outflows in Vul OB1 and IRDC G53.2}
  \label{fig:alpha}
\end{figure} 

Figure \ref{fig:alpha} shows the cumulative distribution function of the jet-bearing YSOs as a function of the estimated spectral index values. The plot shows that approximately 79\% of the driving source candidates for the identified outflows are Class 0/I sources, 17\% are Class II YSOs and 4\% are Class III YSOs. The distribution obtained is consistent with the fact that H$_2$ jets are believed to be most prominent in the early Class 0/I stage and the outflow activity fades away as the YSO evolves. This corroborates with the hydro-dynamical simulations of \citet{Vorobyov_2015}, who show that episodic accretion events, induced by gravitational instabilities, and disk fragmentation are present mostly during the early evolutionary stages of a protostar. 

It should be noted that there could be an inherent observational bias towards embedded Class 0/I YSOs in our study. Due to the molecular gas and dust envelope surrounding Class 0/I YSOs, MHOs are more prominent for early stage protostars. However, outflows from evolved YSOs may not be detected as MHOs since their immediate environment would have less molecular gas. Sensitive optical observations in H$_\alpha$ and [SII] bands are hence essential to get a better insight into this observed trend.

\subsection{H$_2$ emission knots in outflows}\label{knot_gaps}

Chains of H$_2$ emission knots associated to a jet/outflow are indicative of the fact that the driving YSO has undergone episodic ejection \citep[e.g.][]{Bachiller_1996}, likely due to underlying variability in the mass accretion \citep{Scholz_2013,Machida_2019,Takami_2020}. These accretion bursts could originate from disk-instability in the early evolutionary phases of the YSOs \citep[][]{Zakri_2022}.

It is believed that accretion processes, that include both the declining accretion rates and episodic accretion outbursts, could resolve the luminosity problem of protostars \citep[and references therein]{Dunham_2012,Herczeg_2017}. Several H$_2$ surveys, using the UWISH2 database, have observed typical ejection timescales of few kyr \citep[e.g.][]{Ioannidis_2012b,Froebrich_2016,Makin_2018} suggesting an intermediate eruption timescale, between the well-known FU-Ori and EX-Ori type eruptions in low-mass YSOs. The origin of these eruptions, characterised by moderate to large increase in the brightness of the source, is still debated \citep[][]{Audard_2014,Hales_2020,Szegedi_2020}. Brightening $\sim$~5 magnitudes in the V band is seen in FU-Ori types \citep{Hartmann_1985} as compared to 2-4 magnitudes increase in EX-Ori type eruptions \citep{Hales_2020}. Detailed studies on these periodic ejections can enable a better understanding of the mechanisms involved and help in constraining YSO models.

\begin{figure}
  \centering
  \includegraphics[width=0.45\textwidth]{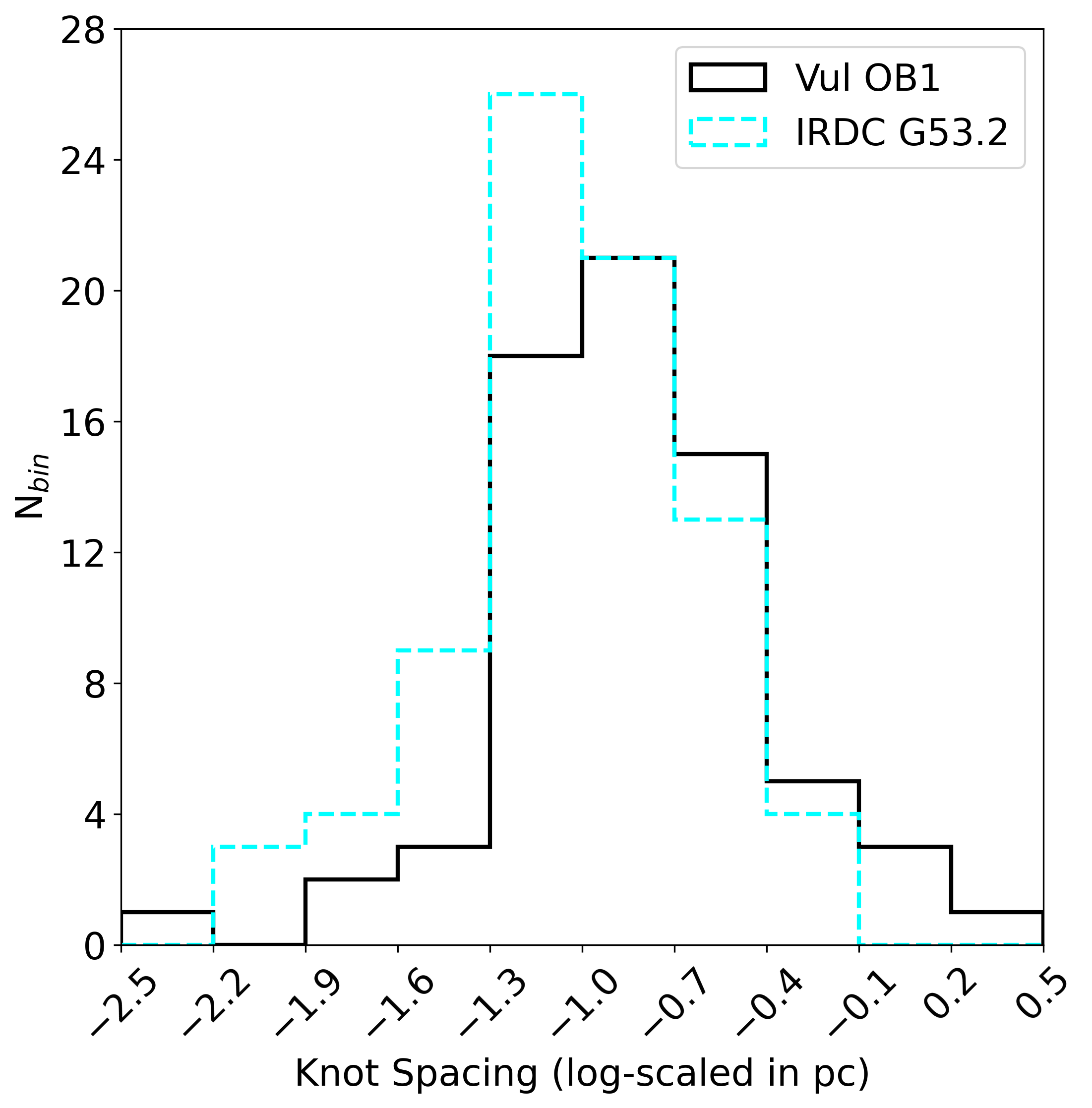}
  \caption{Distribution of knot gaps (in log-scale) for outflows in Vul OB1 and IRDC G53.2.}
  \label{fig:knot_gaps}
\end{figure}

The separation between subsequent H$_2$ emission knots is thus a useful parameter to be determined. This will enable understanding the periodicity in ejection of jets and the possible origin of enhanced accretion in YSOs. These gaps are measured as the separation of mean locations of adjacent H$_2$ knots seen in the continuum-subtracted images. The distribution of knot gaps for MHOs in our study is plotted in Figure \ref{fig:knot_gaps} showing a peak at 0.1~pc. Using a transversal velocity of 80 kms$^{-1}$, the typical velocity used in similar studies based on the UWISH2 survey \citep[e.g., ][]{Ioannidis_2012b}, the knots are found to be separated by $\sim$ 1.2 kyr in time for outflows, similar to the timescales observed by \citet{Makin_2018}. Since the majority of the outflows in this study are driven by Class 0/I YSOs, these timescales are consistent with the estimated burst intervals of $\sim$ 1000 yr for early embedded protostars from the {\it Spitzer} and WISE epochs separated by 6.5 yr \citep[see][]{Fischer_2019}. This also corroborates with the recent findings of \citet{Park_2021}, who suggest that YSOs in earlier evolutionary stages have higher amplitudes of variability with recurrence timescale of $\sim$ 1000 yr.

The estimated timescale of the majority of the outflows in Vul OB1 and IRDC G53.2 lies in between the range of 5-50~kyr for FU-Ori \citep{Scholz_2013} and 1-10~yr for EX-Ori \citep{Aspin_2010} type of eruptions. The observed timescales also lie in between the bi-modal distribution for the outburst timescales observed by \citet{Vorobyov_2018} for isolated and clustered luminosity bursts. This implies that we are probably witnessing ejecta from an intermediate class of eruptive variables similar to MNors \citep{Contreras_2017,Contreras_2019}. More detailed models are required to understand these intermediate eruption timescales and to establish a robust link between the accretion and ejection process based on the observed variability \citep{Nony_2020}.

\subsection{Observed correlation between outflow and driving source parameters}

\begin{figure}
  \centering
  \includegraphics[width=0.45\textwidth]{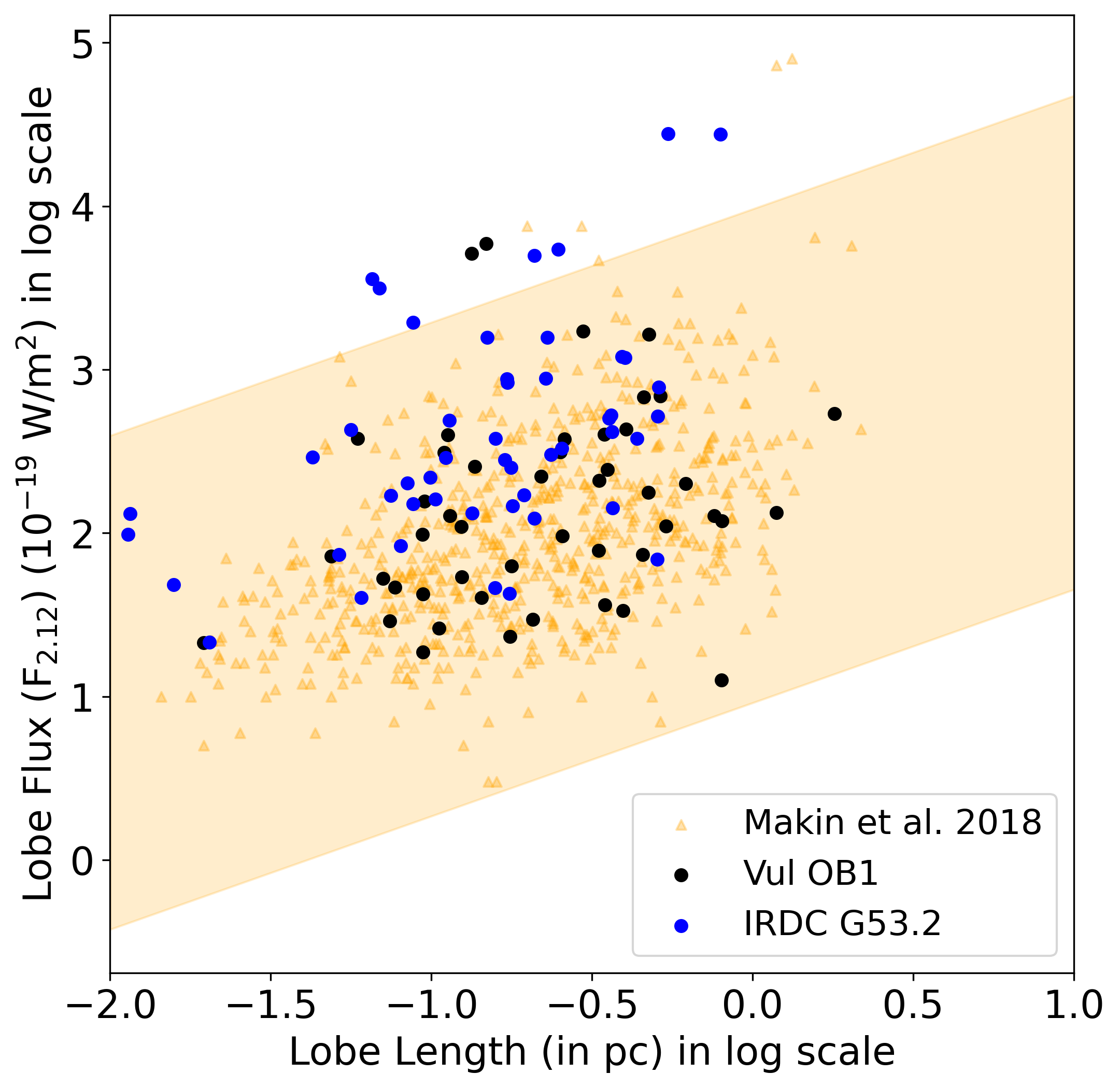}
  \caption[Length of outflow vs outflow H$_2$ 1-0S(1) line flux/luminosity]{Variation of length of outflows identified in Vul OB1 and IRDC G53.2 with the H$_2$ 1-0S(1) line flux densities. The black dots corresponds to outflows identified in Vul OB1, blue dots for IRDC G53.2 and the orange triangles denote outflows from \citet{Makin_2018}. The orange shaded region shows the 3$\sigma$ interval identified using linear regression for outflows from \citet{Makin_2018}.}
  \label{fig:res_length_flux}
\end{figure}

Using the various physical parameters estimated for the outflows and the driving source candidates, we examine the parameter space for trends and correlations, if present. Figure \ref{fig:res_length_flux} represents variation of the outflow lobe length with lobe flux densities. The correlation seen, although weak, indicates that longer outflows are also typically brighter. The figure also plots data from \citet{Makin_2018} and the corresponding 3$\sigma$ interval identified from linear regression between outflow lobe length and flux densities. As seen, the derived parameters of the majority of the outflows identified in Vul OB1 and IRDC G53.2 are consistent with results from the above mentioned study. 

\begin{figure}
  \centering
  \includegraphics[width=0.45\textwidth]{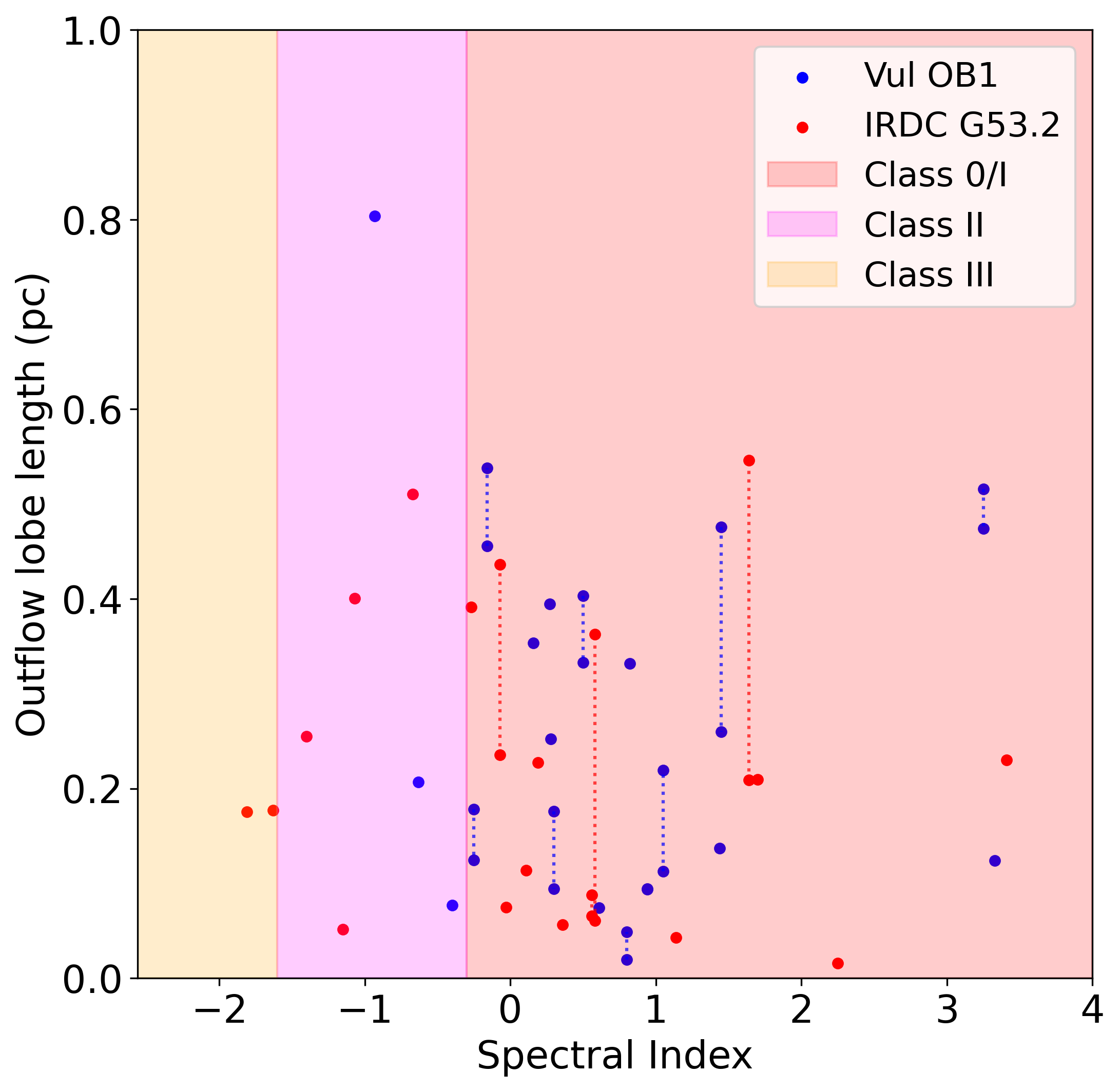}
  \caption{Variation of outflow lobe length with spectral index. The blue and red dots correspond to MHOs in Vul OB1 and IRDC G53.2 respectively. Outflow lobes corresponding to bipolar MHOs are connected using dotted vertical lines.}
  \label{fig:alpha_len}
\end{figure}

A scatter plot of the lobe length of MHOs and the estimated spectral index is shown in Figure \ref{fig:alpha_len}. No correlation between these parameters is evident similar to the findings of \citet{Davis_2009,Zhang_2015}. Inaccurate parameter estimates from H$_2$ observations (e.g. effect of extinction and inclination on lobe length), variation of spectral indices due to variability and confusion with the nearby sources, could affect any correlation, if present. Moreover, lobe lengths are expected to be short at the starting phase of ejection by a YSO. Similarly, jets are expected to fade, thus are more difficult to observe, in the late phase of a YSO with declining accretion.

\begin{figure}
  \centering
  \includegraphics[width=0.45\textwidth]{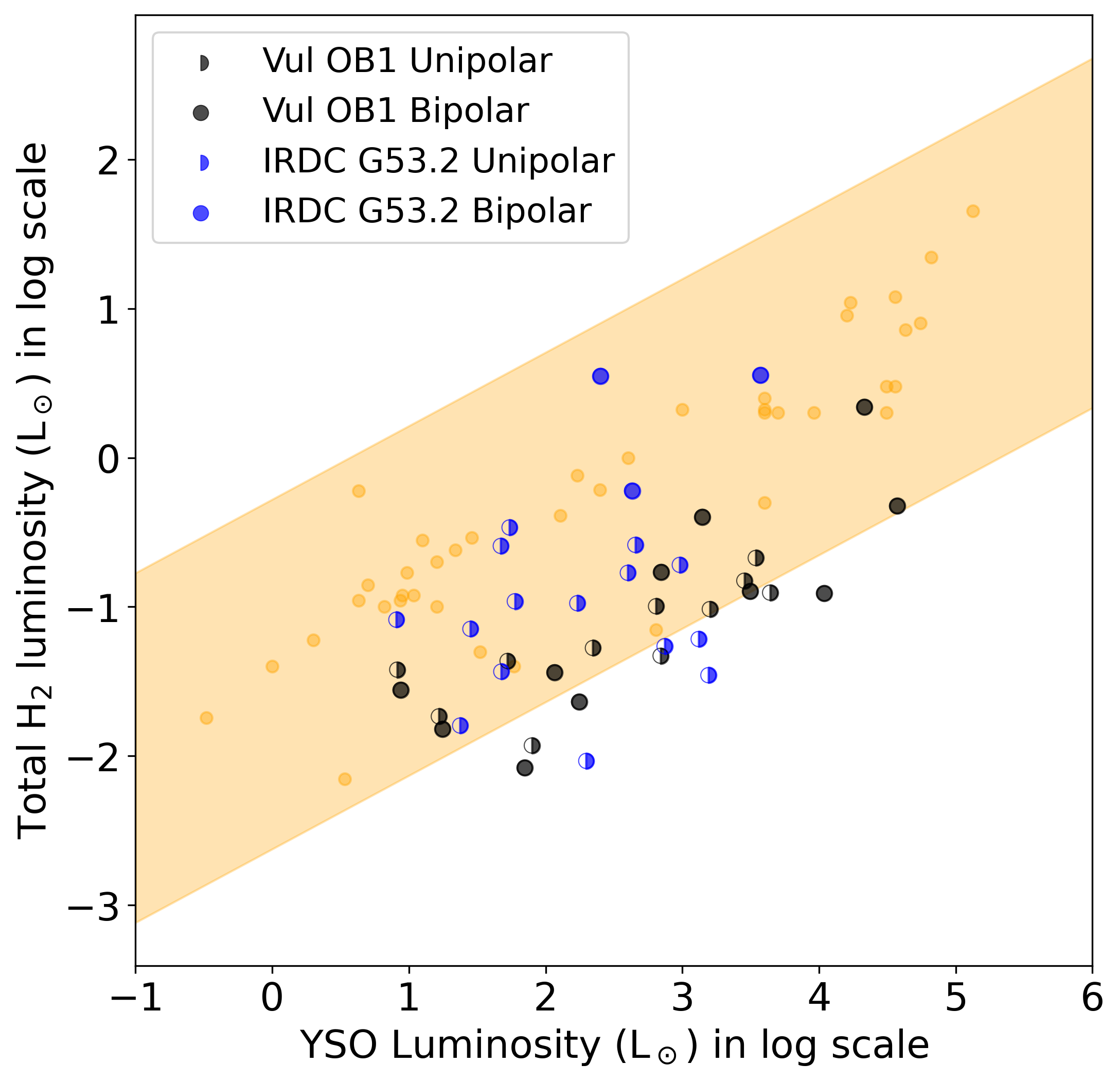}
  \label{fig:res_length_YSOlum}
    \caption{H$_2$ luminosity of the outflows vs photometric luminosity of the driving source candidate. The orange data points are from \citet{Garatti_2006,Garatti_2015} and the corresponding 3$\sigma$ interval is shown by the orange shaded region. All the parameters are in logarithmic scale. The black data points are from Vul OB1 and blue data points are from IRDC G53.2. Bipolar outflows are shown by filled circles, and the half-filled circles represent unipolar flows.}
    \label{fig:YSO_prop_corr}
\end{figure}

In Figure \ref{fig:YSO_prop_corr}, we show the total H$_2$ luminosity of outflows as a function of the photometric luminosity of the corresponding driving source candidates. The figure also shows data points from \citet{Garatti_2006,Garatti_2015}. In agreement with the findings of \citet{Garatti_2006,Garatti_2015}, we observe that MHOs with higher H$_2$ luminosities are driven by more luminous YSOs in our study region. However, we note that the sources in our study deviate from the 3$\sigma$ interval of \citet{Garatti_2006,Garatti_2015}. This deviation could arise from an under-estimation of the H$_2$ luminosities of the outflows as even an uncertainty of 1 magnitude in A$_{\rm V}$ could affect the estimated $L_{2.12}$ by 10\% \citep[see][]{Garatti_2006}. Unlike sources from \citet{Garatti_2006,Garatti_2015}, who have used near-IR spectral line observations to calculate the temperature and extinction, we use a low-resolution extinction map for extinction correction and 2000K as the temperature of the shocked jets. Alternatively, the deviation in the figure could also indicate an over-estimation of luminosity of driving source candidates in our study. Such deviation may be due to lower spatial resolution measurements at longer wavelengths. High resolution mid- and far-infrared observations would be needed to better understand the observed deviation.

Although the correlations observed are not strong, the trends observed pave the way for more detailed studies and confirm the existence of a link between the physical properties of outflows and their driving source candidates. Despite the scatter, these results provide limits on the estimates of various physical properties of both the outflows and the YSOs. These findings paired with high resolution observations can help with our understanding of the origin and dynamics of outflows and YSOs.

\section{Notes on individual outflows}\label{Sec:Notes}

In this work, we have identified 127 jets/outflows displaying interesting morphology and properties. Of the identified outflows, 2 are likely quadrupolar (OF56/MHO2647 and OF57/MHO2646, OF104/MHO2640 and OF105/MHO2641), 2 are associated with dark clouds (OF39/MHO4233 and OF42/MHO4238), 3 outflows show bent or curved morphology (OF1/MHO4218, OF20/MHO4235 and OF33/MHO4244), and 2 are located inside pillars (OF35/MHO4234 and OF37/MHO4249). We find enhanced green emission in the IRAC colour-composite images for 6 outflows implying that they could possibly belong to the population of EGOs. Brief discussion on some of these interesting outflows are given here and discussion on the rest are available in the electronic version. 

\subsubsection*{A protocluster with cluster of jets in IRDC G53.2}

\begin{figure}
  \centering
  \includegraphics[width=0.48\textwidth]{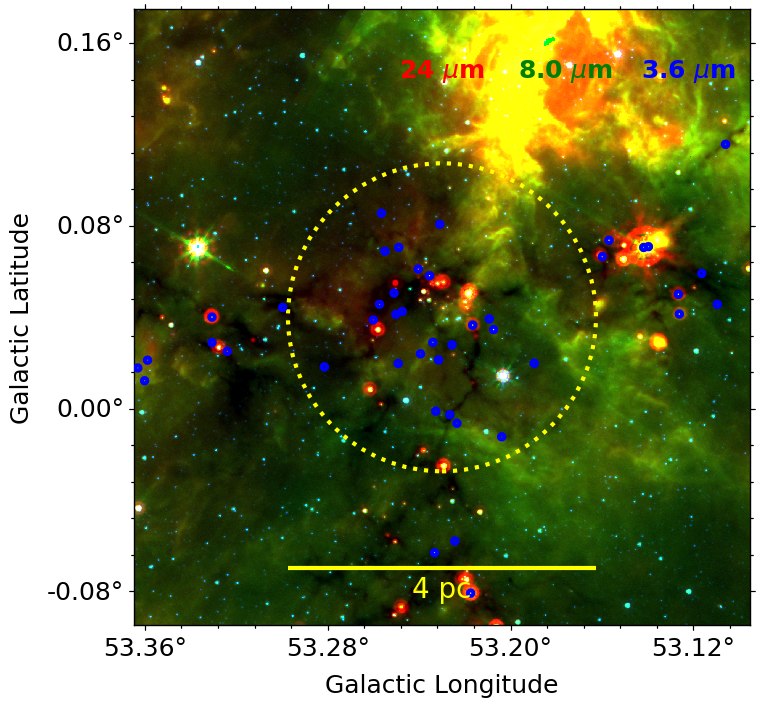}
  \caption{Multi-wavelength color-composite of a hub-filament system in IRDC G53.2 (IRAC 3.6 $\mu$m as blue, 8.0 $\mu$m as green and MIPS 24 $\mu$m as red). MHOs identified in this study are shown by blue dots. The yellow dotted circle indicates the central region of the hub-filament system.}
  \label{fig:hub-filament}
\end{figure}

Figure \ref{fig:hub-filament} shows the IRAC color-composite of an active star forming region (l$\sim$53.23$^\circ$, b$\sim$0.04$^\circ$) within the IRDC G53.2. We observe a clustering of YSOs in a central region with a radius of 2 pc. In the infrared, the region appears to be dark, while from \citet[][]{Wang_2020} we find that it corresponds to a clumpy structure with enhanced emission in dense gas tracers such as HCN, HCO$^{+}$ and N$_2$H$+$. The region also corresponds to an embedded cluster, "G3CC 70", detected in the {\it Spitzer}-GLIMPSE survey by \citet[][]{mora13}. Thus, it is an ideal candidate for studying the role of protostellar feedback on the formation and growth of star clusters \citep[e.g.][]{mur18, rosen20, ver22} using high-resolution molecular observations \citep[e.g.][]{mau09, naka11, bagu21}. 

\begin{figure*}
  \centering
  \includegraphics[width=0.85\textwidth]{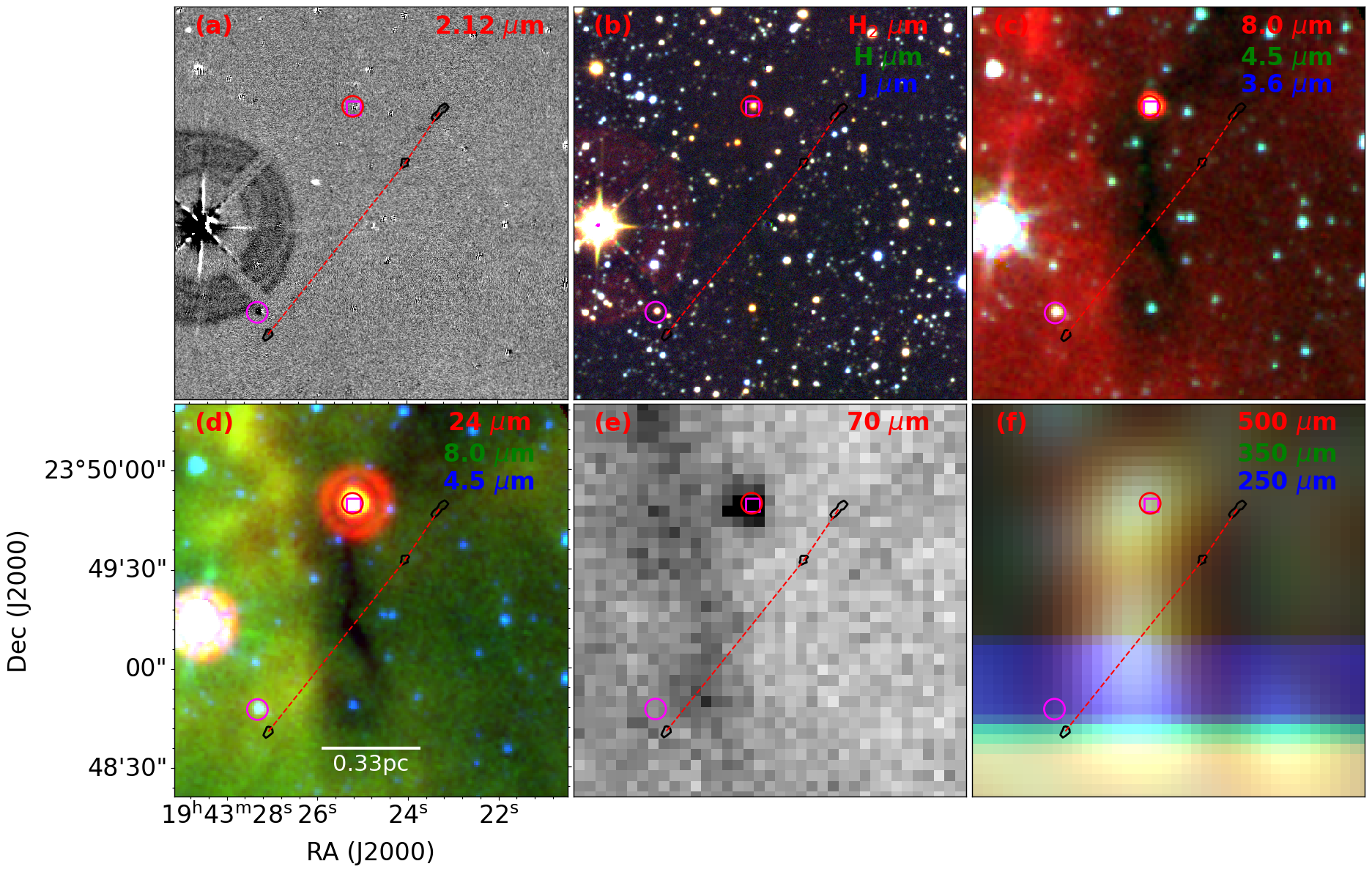}
  \caption{Multi-band image of outflow OF39/MHO4233 originating from a dark cloud. The layout, colours and symbols are the same as Figure \ref{fig:sample_multi}.}
  \label{fig:SC_1}
\end{figure*}

\begin{figure*}  
  \centering
  \includegraphics[width=0.85\textwidth]{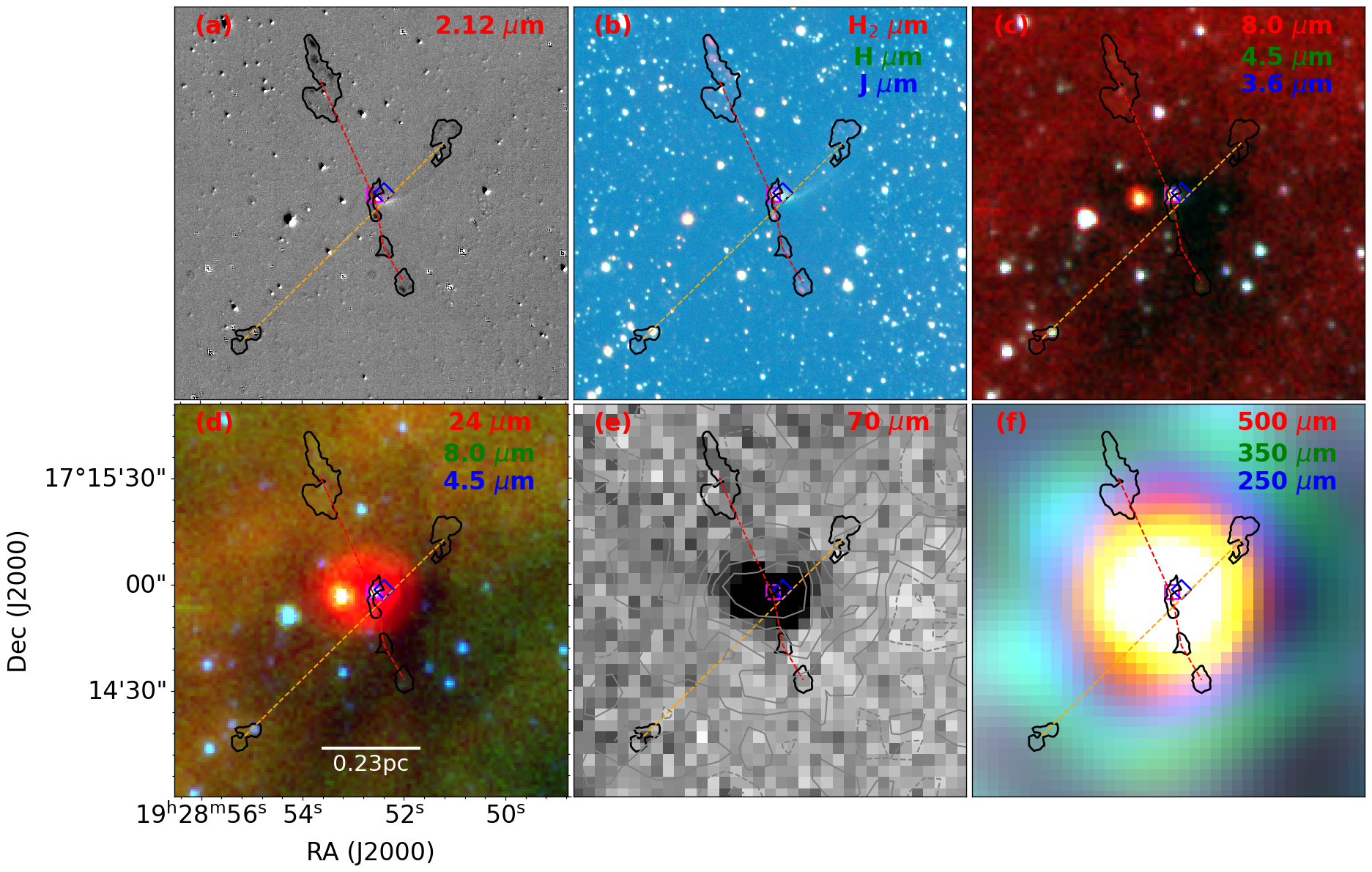}
  \caption{Multi-band image of a quadrupolar outflow OF104/MHO2640, OF105/MHO2641 in IRDC G53.2. The two outflows are shown by red and orange dashed lines. The figure follows the same layout, colours and symbols as Figure \ref{fig:sample_multi}}
  \label{fig:SC_2}
\end{figure*}

\begin{figure*}
  \centering
  \includegraphics[width=0.85\textwidth]{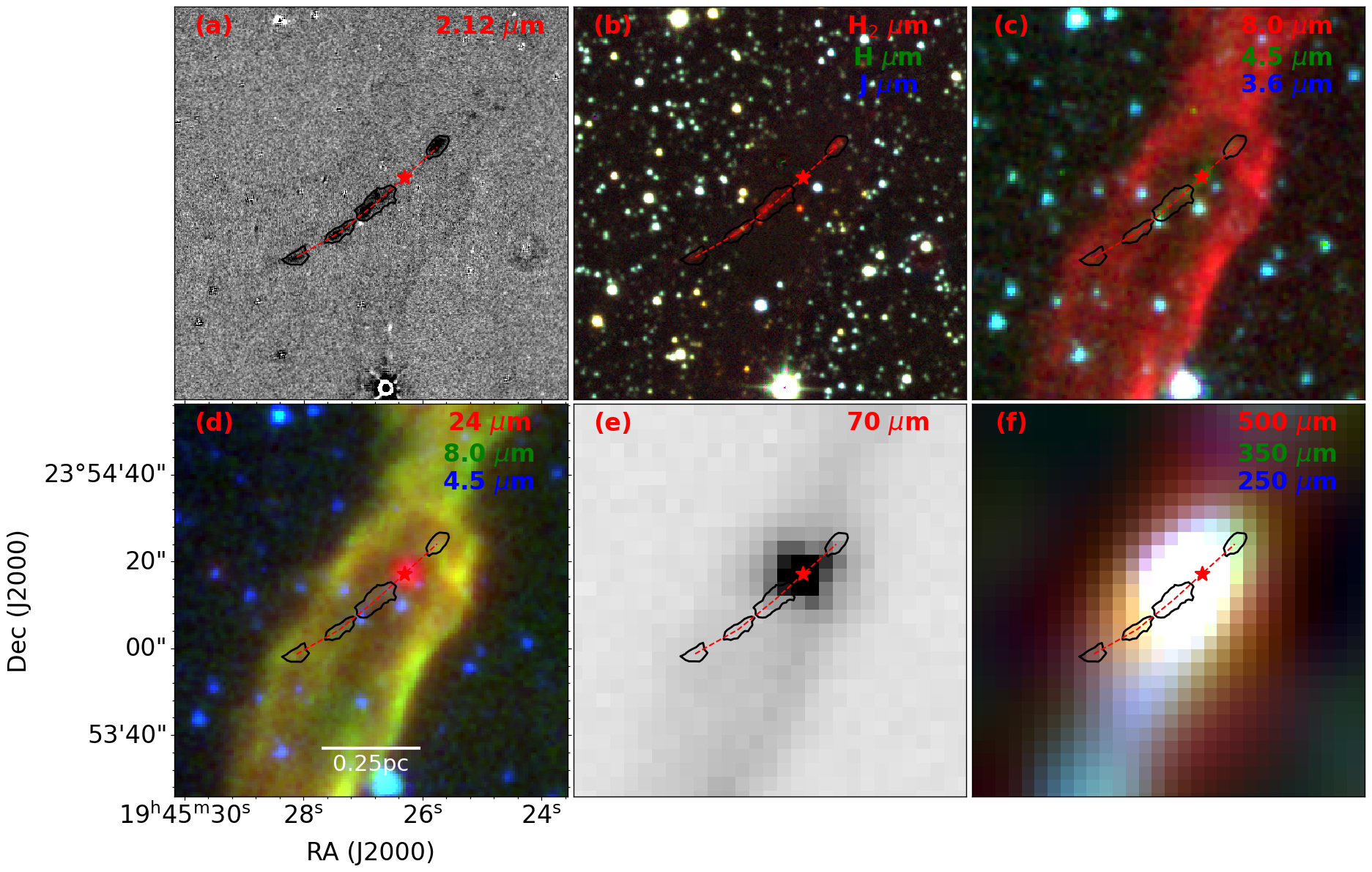}
  \caption{Multi-band image of outflow OF37/MHO4249 inside a pillar in Vul OB1. The layout, colours and symbols are the same as given in Figure \ref{fig:sample_multi}.}
  \label{fig:SC_3}
\end{figure*}

\subsubsection*{OF33/MHO4244}

Figure \ref{fig:sample_multi} shows multi-band colour composite image of a curved outflow identified in the Vul OB1. The C-shaped geometry depicting the curvature of the outflow can arise from the motion of the surrounding ambient medium with respect to the outflow or from high proper motion of the YSO itself \citep{Bally_2007, Cunningham_2009, Reipurth_2001}. Similarly, S-shaped geometries have also been found in literature \citep[e.g. see HH 315 in][]{Reipurth_1997}, which suggests a precessing jet likely due to the presence of a companion to the outflow driving source. 

\subsubsection*{OF39/MHO4233}
Figure \ref{fig:SC_1} shows multi-band colour-composite images of the identified outflow OF39/MHO4233 in the Vul OB1 complex and the region associated with it. The figure displays an interesting case of an outflow likely originating from a deeply embedded YSO inside the dark cloud. No driving source is identifiable up to 70 $\mu$m. The outflow is observed to be oriented away from the axis of the filamentary dark cloud. Such alignments between outflows and filaments could be a result of continuous transport of mass from the filament onto the protostellar core in the presence of moderately strong magnetic field lines \citep[e.g.][and reference therein]{Kong_2019}. These deeply embedded candidates are ideal targets to study the earliest stages of protostellar evolution.

\subsubsection*{OF104/MHO2640 and OF105/MHO2641}
In Figure \ref{fig:SC_2}, we show the multi-band images of a likely quadrupolar outflow, OF104/MHO2640  and OF105/MHO2641, identified in the IRDC G53.2 comprised of two bipolar outflows. The driving source candidates responsible for the two outflows are not resolved in the images presented. Better sensitivity and higher resolution images are required to detect the YSOs responsible for individual outflows. Similar multi-polar outflows, such as HH 124 and HH 571/572 from NGC 2264N and HH 111, have been observed and are of great importance as they could be potential indicators of binary star formation \citep{Reipurth_1999, Reipurth_2004, Makin_2018, Chase_2017}.

\subsubsection*{OF37/MHO4249}
We discover an interesting case of an outflow, OF37/MHO4249, inside one of the pillars in Vul OB1 \citep[named VulP1 in][]{Billot_2010}. The corresponding multi-band images for this 0.5 pc long outflow are shown in Figure \ref{fig:SC_3}. The driving source candidate of the outflow is a PBR implying an early Class-0 object. Though enhanced green emission is discernible, the source has not been identified in existing catalogs of EGOs \citep{Cyganowski_2008,Chambers_2009}. The luminosity of the source is estimated to be $\sim$ 75 L$_\odot$ using the correlation between the 70 $\mu$m flux and bolometric luminosity of protostars as observed by \citet{Dunham_2008} as the source lacked sufficient data points for SED-fitting. This outflow presents itself as an excellent site of triggered star formation and is an ideal target for follow-up studies. Another case of triggered star formation inside a pillar has been discussed in the online material (see OF35/MHO4234).

\section{Conclusions}\label{Sec:Conclusion}

We present the result of a statistical analysis of jet bearing YSOs in two interesting star forming complexes, the Vul OB1 association and the IRDC G53.2. We probe 12 square degrees of archival data from the UWISH2 survey and identify a total of 127 outflows in these two regions. 
Using photometric data from multiple Galactic surveys and existing YSO catalogs from the literature, the driving source candidates are identified for 79 outflows (62 \% of the total number of outflows identified in the two clouds). The CO velocity-integrated maps for the two clouds show that most of the MHOs identified lie within the cloud boundary, hence, mostly associated with it.

Various estimates of several physical parameters of the MHOs and their driving source candidates have been presented. Physical parameters of 45 driving source candidates have been derived using SED modelling.

The important results obtained from our analysis are summarised below:
\begin{itemize}
  \item The outflows associated with the two clouds are seen to be predominantly driven by intermediate-mass protostars.
  
  \item Outflows have a typical ejection frequency of 1.2 kyr for a typical transversal velocity of 80 kms$^{-1}$.
  
  \item We study and identify various correlations between the physical parameters of the outflow with the physical parameters of the candidate driving source. Despite the scatter observed in our plots, clear trends are seen from the analysis suggesting that brighter YSOs drive more luminous outflows.
  
  \item The H$_2$ 1-0S(1) line flux of outflows is found to be correlated with the length of the outflows. Our results are in agreement with the outflow population of other Galactic plane studies using UWISH2. 
  
  \item Several interesting star forming regions have been identified based on the spatial distribution of outflows. 
  
  Many outflows displaying extended green emission in the IRAC bands, and hence likely EGOs, are identified within the two clouds. These are ideal test laboratories to study massive star formation. We also identify two jets which are spatially located inside the pillars seen in the Vul OB1 complex. One of these outflows inside the pillars is of particular interest and is also characterised by excess extended emission at 4.5 $\mu$m along the axis of the outflow. The outflow is driven by a PBR which is an early Class 0 source and is an ideal target to study possible triggered star formation inside a pillar.
  
\end{itemize}

\noindent
Although we discuss various trends and correlations between physical parameters of outflows and their driving source candidates, the scope of our analysis is restricted by the limited data available for the regions. We also acknowledge that, given the lower statistics of outflows driven by Class II and Class III YSOs, the observed trends may be mainly representative of Class 0/I YSOs. Our analysis represents an ideal sample for multi-wavelength analysis employing data from existing as well as upcoming astronomical facilities. Detailed analysis of the special cases discussed will provide a deeper insight into the star formation process and provide clues to address crucial aspects of star formation.

\section*{Data Availability}\label{Sec:Data_availability}
The reduced continuum subtracted H$_2$ images used in this work are publicly available at the UWISH2 survey archive: \url{http://astro.kent.ac.uk/uwish2/index.html}. The UKIDSS near-infrared images are publicly available at WSA - WFCAM Science Archive: \url{http://wsa.roe.ac.uk/}. The Herschel and Spitzer images are publicly available at NASA/IPAC Infrared Science Archive: \url{https://irsa.ipac.caltech.edu/frontpage/}. All the point source catalog data are publicly available on Vizier. Multiwavelength cutouts of the individual MHOs can be obtained upon reasonable request to the authors.


\bibliographystyle{mnras}
\bibliography{Bib_file} 


\section*{Supporting Information}\label{Sec:Supplementary}
Additional Supporting Information may be found in the online version of this article:

APPENDIX A: Full list of MHO and YSO parameters as provided in Table \ref{Tab:Out_prop} and Table \ref{Tab:YSO_prop}

APPENDIX B: Images of MHOs identified in the study region. 

\bsp	
\label{lastpage}
\end{document}